\documentclass[letterpaper,twocolumn,10pt]{article}
\usepackage{usenix}

\usepackage[normalem]{ulem}
\usepackage{graphicx}
\usepackage{xcolor}
\usepackage{soul}
\usepackage{adjustbox}
\usepackage{caption}
\usepackage{subcaption}
\usepackage{booktabs}
\usepackage{multirow}
\usepackage{url}
\usepackage{tabularx}
\usepackage{cleveref}
\usepackage{wrapfig}
\usepackage{balance}
\usepackage{etoolbox}
\usepackage{enumitem}
\usepackage{xspace}
\usepackage[binary-units=true]{siunitx}
\newcommand{\SIx}[1]{\num{#1}\relax}
\DeclareSIUnit\per{/}
\DeclareSIUnit\packet{packet}

\makeatletter
\def\SOUL@hlpreamble{%
\setul{}{2.2ex}
\let\SOUL@stcolor\SOUL@hlcolor
\SOUL@stpreamble
}
\makeatother
\soulregister\st7
\soulregister\SIx7
\soulregister\SI7
\soulregister\cite7
\soulregister\ref7
\soulregister\cref7
\soulregister\Cref7
\soulregister\eg7
\soulregister\ie7
\soulregister\texttt7
\soulregister\xspace7
\soulregister\etal7

\usepackage{array}
\newcolumntype{L}[1]{>{\raggedright\let\newline\\\arraybackslash\hspace{0pt}}m{#1}}
\newcolumntype{C}[1]{>{\centering\let\newline\\\arraybackslash\hspace{0pt}}m{#1}}
\newcolumntype{R}[1]{>{\raggedleft\let\newline\\\arraybackslash\hspace{0pt}}m{#1}}

\usepackage{tikz}
\usetikzlibrary{external}
\usetikzlibrary{patterns}
\usepackage{pgfplots}
\usepackage{pgfplotstable}
\usepgfplotslibrary{fillbetween}
\usetikzlibrary{pgfplots.groupplots}
\usetikzlibrary{arrows}
\usetikzlibrary{patterns}
\usetikzlibrary{positioning}
\usetikzlibrary{decorations.pathreplacing}
\usetikzlibrary{shapes.arrows}
\usetikzlibrary{shapes.geometric,shapes.misc}
\usetikzlibrary{pgfplots.groupplots}
\pgfplotsset{compat=newest}
\pgfkeys{/pgf/number format/.cd,1000 sep={}}

\pgfplotsset{
    discard if/.style 2 args={
        x filter/.code={
            \edef\tempa{\thisrow{#1}}
            \edef\tempb{#2}
            \ifx\tempa\tempb
                
            \fi
        }
    },
    discard if not/.style 2 args={
        x filter/.code={
            \edef\tempa{\thisrow{#1}}
            \edef\tempb{#2}
            \ifx\tempa\tempb
            \else
                
            \fi
        }
    }
}

\newcommand{\eg}{e.g.,\xspace}
\newcommand{\ie}{i.e.,\xspace}
\newcommand{\etal}{et~al.\xspace}

\usepackage{pifont}
\newcommand{\circled}[1]{\ding{\numexpr#1+191}}
\newcommand{\circledtext}[1]{\raisebox{.5pt}{\textcircled{\raisebox{-0.9pt} {\footnotesize #1}}}}

\begin{document}

\date{}

\title{Side-Channel Attacks on Open vSwitch}

\author{
{\rm Daewoo Kim}\\
University of Waterloo
\and
{\rm Sihang Liu}\\
University of Waterloo
}

\maketitle

\begin{abstract}
    
Virtualization is widely adopted in cloud systems to manage resource sharing among users.
A virtualized environment usually deploys a virtual switch within the host system to enable virtual machines to communicate with each other and with the physical network. 
The Open vSwitch (OVS) is one of the most popular software-based virtual switches.
%
%
It maintains a cache hierarchy to accelerate packet forwarding from the host to virtual machines. 
We characterize the caching system inside OVS from a security perspective and identify three attack primitives.
Based on the attack primitives, we present three remote attacks via OVS, breaking the isolation in virtualized environments.
First, we identify remote covert channels using different caches.
Second, we present a novel header recovery attack that leaks a remote user's packet header fields, breaking the confidentiality guarantees from the system. 
Third, we demonstrate a remote packet rate monitoring attack that recovers the packet rate of a remote victim.
To defend against these attacks, we also discuss and evaluate mitigation solutions.
\end{abstract}



\section{Introduction}

Side-channel attack is a category of attacks that exploits extra information from unintended channels rather than vulnerabilities in software or algorithms.
In computer systems, side-channel attacks threaten and intercept valuable information through both hardware and software. 
Caches are common targets of side-channel attacks, as their hit/miss timing can be used to infer secret information about the program, bypassing the existing system-level isolation and protection. 
For example, CPU caches \cite{liu2015last, yarom2014flush+, kayaalp2016high, gullasch2011cache, guo2022adversarial} and TLBs \cite{gras2018translation} can be used to infer program behavior and even data values executed by the program; database caches \cite{shahverdi2021database, dbreach2023hogan} leak database queries and can be used to construct the database.
To perform side-channel attacks on caches, the attacker typically initializes the cache to a known state and then observes any changes in the state made by the target application that shares the same cache. 
For instance, Prime+Probe \cite{liu2015last} and Flush+Reload \cite{yarom2014flush+} are common approaches. 

As modern systems are being moved to the cloud for better efficiency and maintenance, resource sharing becomes more common. 
At the same time, such sharing scenarios are more susceptible to side-channel attacks, as different applications and services can be co-located on the same server, sharing not only hardware components but also part of the software stack. 
Prior works have demonstrated side-channel attacks in remote cloud scenarios. 
For example, NetSpectre \cite{schwarz2019netspectre} proposes a remote Spectre attack that reads memory over the network using Evict+Reload, and NetCat~\cite{kurth2020netcat} exploits the Data-Direct I/O (DDIO) to infer the last-level cache hit/miss over the network using Prime+Probe.
These remote side-channel attacks are stealthy as an attacker can leak secret data over the network. 


In a cloud environment, virtualization is a common approach to allow multiple users to share the same server, where the hypervisor manages a number of virtual machines (VMs).
The network that connects these VMs is usually also managed by network virtualization systems that apply network policies to incoming packets and use software switches to forward packets to different VMs by bridging them between the physical network interface and the VMs. 
The Open vSwitch (OVS) is one of the most widely used software switches \cite{pfaff2015ovs,tu2021ovsten} that follows the OpenFlow standard \cite{mckeown2008openflow}.
Based on the packet flow that travels from a source to a destination, OpenFlow specifies actions that an incoming packet will take (such as being forwarded to a certain port) by looking up a number of OpenFlow tables in sequence.
OVS accelerates the sequential table lookup process with two levels of software-based caches --- a fast but small \emph{microflow cache} (with a default size of 8192 as of OVS v3.6 with the DPDK datapath) and a slower but larger \emph{megaflow cache}.
The caches reside in the main memory, containing structures that match the network packet header and actions that the matching packets will take.
\emph{Hit} in either cache is referred to as the \emph{fast path}.
An incoming packet first looks up the microflow cache. 
A microflow miss will be directed to the megaflow cache. 
Eventually, megaflow cache misses will perform the slow OpenFlow table lookup, which is referred to as the \emph{slow path}. 
In summary, OVS is a complex software system with caching, intended to be shared among VMs, and can be accessed remotely via network.

In this work, we aim to examine and assess the security implications of OVS --- whether attackers can violate the confidentiality guarantees in the virtualization system and obtain unauthorized access to secret information from remote users through OVS.
There have been prior security studies on OVS systems. However, they only focus on performance degradation due to malicious users that can lead to denial-of-service (DoS) \cite{Csikor2020TupleSpaceOVS, CVE-2023-3966, CVE-2020-35498, CVE-2019-25076, CVE-2017-9263}.
However, the side channel aspect largely remains unstudied. 

To exploit side channels in OVS, a key prerequisite is a good understanding of its characteristics and behaviors.
Although OVS is open-sourced, its security implications remain unclear.
Therefore, we characterize OVS from a security angle, by evaluating latencies, eviction and timeout policies, and internal structures of each cache. And then, we cross-reference our findings with the OVS documentation.
We summarize our findings as three attack primitives. 
(1)~Each OVS component has distinguishable latencies: microflow cache, megaflow cache, and slow path in OpenFlow feature round-trip times (RTTs) of \SI{270.08}{\micro\second} (with default microflow cache size), \SI{277.18}{\micro\second}, and \SI{1196.03}{\micro\second}, respectively, in a one-hop network system. 
The latency differences can be exploited to infer cache states.
(2)~The microflow cache evicts an existing flow entry upon a collision with the hash of a new packet header.
As the hash value is generated from five header fields (IP addresses and port numbers of source and destination, and protocol), the microflow cache can leak the packet header fields.
(3)~The megaflow cache keeps a number of subtables for incoming packets to look up and determine their actions.
The lookup is sequential and completes upon the first hit, and frequently accessed subtables are ordered at the front.
Thus, the access frequency controls the subtable ordering and affects latency, which can be leveraged to infer packet rate. 
Next, we present three attacks using these attack primitives.

First, we establish and demonstrate remote covert channels using both the microflow cache and the megaflow cache.
Such covert channels involve a remote sender and receiver who do not have direct communication.
Instead, the sender has access to a service (Memcached in our experiment) that is co-located with the receiver.
The service and the receiver are running in separate VMs but share an OVS. 
To send secret data, the remote sender queries the service to generate contention on the OVS.
In parallel, the receiver queries another service located in a separate server to sense this contention. 
The microflow-cache-based covert channel leverages the latency difference between microflow cache hit and miss.
Depending on the bit that the sender transmits, the sender either generates a collision on a specific microflow entry or not, which affects the receiver's RTT and can be detected. 
The megaflow-cache-based covert channel exploits the latency difference among subtables and their reordering mechanism in the megaflow cache.
The sender controls the packet rates to order the target subtable at the desired location.
The location can then be detected by the receiver based on the latency difference, thus retrieving the bit. 
Covert channels based on the microflow and the megaflow caches achieve \SI{15.8}{\bit\per\second} and \SI{0.73}{\bit\per\second}, respectively, with low error.

Second, we demonstrate an attack that retrieves a remote user's IP address and port number via the microflow cache.
We also assume that the attacker can query a service that is co-located with the victim and shares the OVS. 
We assume that the victim is accessing a publicly accessible service that is known by the attacker.
Thus, among the five packet header fields that generate a hash for the microflow cache index, the destination IP + port, and the protocol are known. 
The attacker only needs to recover the victim’s source IP + port based on hash collisions.
The attacker performs a Prime+Probe attack to identify entries that are evicted by the victim, by measuring the RTTs of all microflow cache entries via queries to the co-located service. 
Then, the attacker computes the hashes of all possible victim's IP and port combinations to find the one that matches the hash values of the evicted entries.
By probing the victim for \SI{7}{\minute} (with default microflow cache size), the attacker can successfully recover these fields at an accuracy of 91\,\%, breaking the isolation among virtual machines.

Third, we present a packet rate monitoring attack, where an attacker infers the packet rate of a remote user.
We assume that a similar scenario as the packet header recovery attack, where the attacker can query a publicly accessible service that is co-located with the victim and shares the OVS. 
The victim sends packets to another remote service that is known to the attacker, \eg can be detected using the packet header recovery attack.
The attacker exploits the packet-rate-dependent reordering of the megaflow cache subtables to recover the victim's packet rate.
The attacker first evicts the victim's flow in the microflow cache, which results in the increment of the megaflow subtable hit counts corresponding to the victim's traffic.
In parallel, the attacker orders other subtables by accessing them at a known rate, and probes the victim's subtable. 
The relative latency difference between the other subtables and the victim-accessed subtable reveals the relative packet rate difference. 
This way, the attacker can track the range of the victim's packet rate in real time. 
Our evaluation with real-world network traces demonstrates a 71.92\,\% monitoring success rate (with default microflow cache size).

Finally, we propose and evaluate three mitigation mechanisms for OVS side channels: isolation of OVS instances, hash randomization for the microflow cache, and subtable reordering randomization for the megaflow cache. 
Our evaluation shows that these methods effectively mitigate the side channels we identified in this work.

The contributions are summarized as follows: 
\begin{itemize}[leftmargin=*, nolistsep]
    \item We thoroughly study and characterize caching in OVS from the security aspect.
    \item We establish remote covert channels using both microflow cache and megaflow cache in OVS.
    \item We present a novel remote packet header recovery attack that retrieves a remote user's IP address and port number from network packets.
    \item We demonstrate a novel remote packet rate monitoring attack that identifies a remote user's real-time packet rate. 
    \item We propose and evaluate three defense mechanisms for OVS, demonstrating effective mitigation of side channels.

\end{itemize}
\section{Background} \label{sec:background}

In this section, we first introduce side-channel attacks, and then OpenFlow and Open vSwitch (OVS). 

\subsection{Side-Channel Attacks} \label{subsec:sidechannel}
Side channels are communication channels based on indirect, unintended behaviors and features. 
In computer systems, side channels widely exist.
Attackers can perform side-channel attacks by leveraging these channels to secretly leak information about the target system.
For example, an attacker can exploit power \cite{randolph2020power,wei2018know,zhao2018fpga} and thermal signals \cite{thermalbleed,Garg2023ThermWare} of computing systems to bypass the original protections and isolation. 
A secret sender may also transmit information using such unintended channels (\ie covert channels).

One of the major categories of side-channel attacks exploits timing differences of caches in computer systems.
Caching is a common technique to provide fast access to commonly used data, such as caches in CPUs and databases. 
Accessing data in a cache (\ie cache hit) has lower latency than those not in the cache (\ie cache miss). 
An attacker can infer secret information about programs using such timing differences.
There are several common methods to perform side-channel attacks on caches. 
For example, Prime+Probe~\cite{liu2015last} leaks information about whether a victim uses a certain cache set.
First, the attacker fills a cache set with its data and then measures the time to read the data.
If the victim has accessed the same cache set, the attacker’s data is evicted, causing longer reaccess time.
Flush+Reload~\cite{yarom2014flush+} exploits shared memory (\eg a shared library) and leaks information about whether a victim program loads specific data into the cache.
The attacker flushes a targeted memory location and measures the reload time.
If the victim has loaded the corresponding cache line into the cache, the reload time becomes shorter.

Even more stealthily, cache-based side-channel attacks utilizing these techniques can be executed over a network.
Examples include NetSpectre~\cite{schwarz2019netspectre} and NetCAT~\cite{kurth2020netcat}, which demonstrated that the caching timing differences are significant enough to be discerned even after traversing a network.
As cloud systems commonly share computing platforms over the network, the capability of launching side-channel attacks over the network is a prominent threat. 

Although side channels widely exist and are hard to detect, performing a side-channel attack has a key prerequisite, that is good knowledge about the target system.
For example, performing a side-channel attack on the CPU cache requires reverse-engineering of CPU cache structure (\eg size and associativity) and its latency \cite{gullasch2011cache, guo2022adversarial, kayaalp2016high, liu2015last, yarom2014flush+};
one that targets a database system would need full knowledge about the database algorithm and software implementation \cite{shahverdi2021database,dbreach2023hogan}.

\subsection{OpenFlow and Open vSwitch (OVS)} \label{subsec:ovs}

\begin{figure}[t]
  \centering
  \includegraphics[width=1\linewidth]{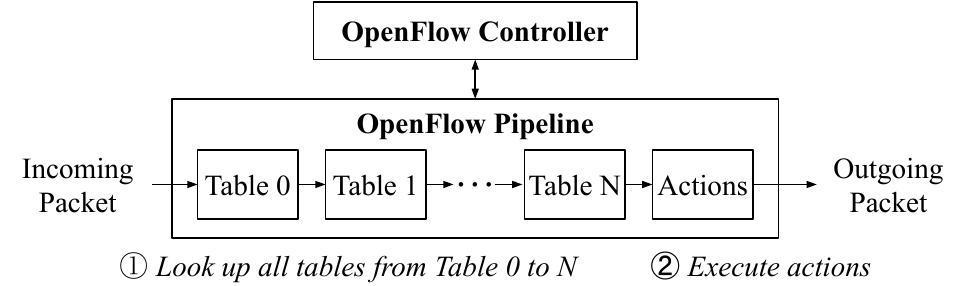}
  \caption{Packet forwarding in OpenFlow pipeline.}
  \label{fig:OpenFlow_diagram}
\end{figure}

Conventionally, network devices have fixed functions to control network traffic.
Therefore, each function in a device has to be separately maintained to control network traffic.
In this section, we introduce OpenFlow and Open vSwitch which alleviate the burden of network management.  

\subsubsection{OpenFlow}
OpenFlow~\cite{mckeown2008openflow} is a software-defined networking (SDN) control protocol. 
\Cref{fig:OpenFlow_diagram} demonstrates how a switch following OpenFlow protocol works.
The OpenFlow controller is a centralized unit that decides the routes of packet flows to the final destinations.
A packet flows from a source to a destination via the route decided by the controller in OpenFlow protocol.
Flow tables decide the \emph{action} that incoming packets will take by \emph{matching} packet header fields with each table.
If the requested flow information does not exist, the OpenFlow controller updates the flow tables based on the user-specified OpenFlow rules. 
\Cref{fig:OpenFlow_diagram} demonstrates the OpenFlow pipeline, where an incoming packet goes through tables (Table 0 to N) in order (step \circled{1}). Eventually, the packet takes actions based on the table lookup at the end of the pipeline (step \circled{2}), such as forwarding the packet to a certain port.

\begin{figure}[t]
  \centering
  \includegraphics[width=1\linewidth]{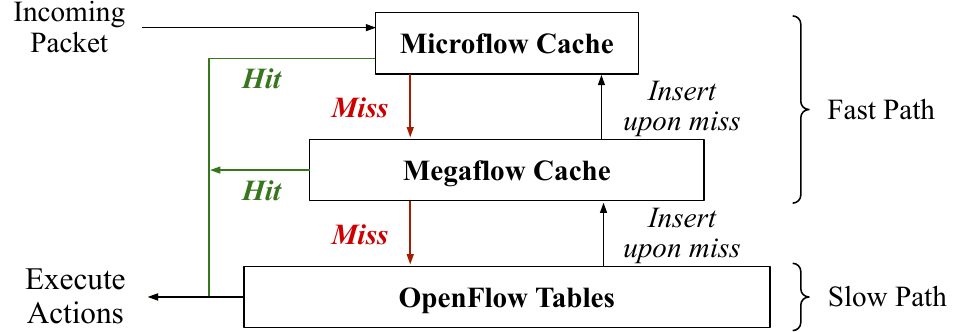}
  \caption{Software-based caches in OVS.}
  \label{fig:Cache_diagram}
\end{figure}

\begin{figure*}
  \includegraphics[width=\linewidth]{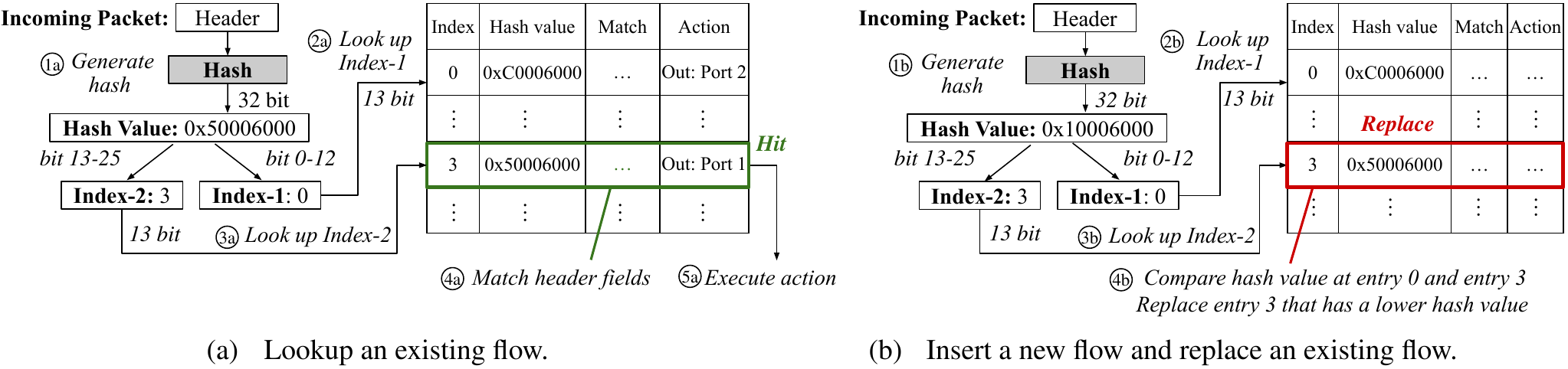}
  \caption{Microflow cache (with default size of 8k entries). }
  \label{fig:Microflow_diagram}
\end{figure*}

\subsubsection{Open vSwitch}
The Open vSwitch (OVS) \cite{pfaff2015ovs,tu2021ovsten} is one of the most widely-used virtual switches based on the OpenFlow protocol.
Like physical switches that follow the OpenFlow protocol and go through table lookups, OVS does the same for virtualization environments, being managed by the OpenFlow controller. 
In virtual environments, OVS establishes connections among virtual machines (VMs) and the physical NIC on the host.

The conventional OpenFlow table lookups always happen sequentially for all incoming packets, slowing down packet forwarding. 
To accelerate the sequential OpenFlow table lookup process, OVS maintains two levels of caches. 
These caches are implemented in the software and kept in the main memory. 
\Cref{fig:Cache_diagram} shows the cache hierarchy inside OVS, where the \emph{microflow cache} is the first structure to be accessed upon an incoming packet, followed by the \emph{megaflow cache} if the microflow cache misses. Accesses within the caches are referred to as the \emph{fast path}. 
Only when both caches miss, the packet is directed to the OpenFlow tables, which is the \emph{slow path}. 
Much like caching in CPUs, once the packet is handled by a lower-level cache in the hierarchy, its corresponding flow will be inserted into the upper levels of caches. 


\subsection{Caching in Open vSwitch} \label{subsec:caching}


OVS contains a small but fast \emph{microflow cache} for exact matching and a larger but slower \emph{megaflow cache} that uses wildcard rules for matching, organized into a number of subtables.
Next, we describe both caches in detail. 


\subsubsection{Microflow Cache} \label{subsubsec:microflow}
The microflow cache is the first level of cache in the hierarchy of OVS, as depicted in \Cref{fig:Cache_diagram}.
The microflow cache has a fixed-size table with \SIx{8192} (\ie~8k) entries by default (as of OVS v2.17.9).
Each entry contains one flow. 
The incoming packet needs to be an exact match with the entry for a cache hit, \ie all packet header fields
need to be matched, according to the ``match'' field. 
If an incoming packet hits the microflow cache, it will then take the ``action'' specified in the entry. 
By performing only a single, exact hash lookup, the microflow cache is optimized for low latency but with limited capacity.

\Cref{fig:Microflow_diagram}a demonstrates the lookup process in the microflow cache when it has the default size of 8k entries. 
The microflow cache first generates a 32-bit hash value from five fields (source IP and port, destination IP and port, and protocol) in the packet header (step~\circledtext{1a}).
The 32-bit hash value is then divided into two hash values, forming two 13-bit hash values for two indices in the default 8k-entry microflow cache: Index-1 using bits 0--12 and Index-2 using bits 13--25 (the remaining higher-order bits are unused). 
It first looks up Index-1 which corresponds to entry 0 and compares the full 32-bit hash with the entry (step~\circledtext{2a}).
In this example, entry 0 does not match. Then, the microflow cache looks up Index-2 which corresponds to entry 3 (step~\circledtext{3a})
and finds a matching entry (step~\circledtext{4a}).
Thus, the packet is forwarded to output port 1 according to the action field (step~\circledtext{5a}).
In case of no matching entry in the microflow cache (\ie a miss), OVS forwards the packet to the next-level megaflow cache.
Once a miss is resolved, OVS inserts this flow into the microflow cache.
With this mechanism, increasing the number of bits in each hash index enlarges the microflow cache. For example, a microflow cache with \SIx{16384} (\ie~16k) entries has 14 bits for each index.

According to OVS documentation \cite{pfaff2015ovs,tu2021ovsten}, a flow stays in the microflow cache until it is evicted by a new flow. 
\Cref{fig:Microflow_diagram}b illustrates the eviction process when inserting a new flow. 
Similar to the microflow cache lookup, the hash function generates a 32-bit hash value using the packet header (step~\circledtext{1b}).
Then, it tries to insert the new entry into Index-1 if the entry is available; otherwise, it attempts Index-2. 
If neither entry is available, the microflow cache compares the hash values of both entries and replaces the one with a smaller 32-bit hash value. 
In the example of \Cref{fig:Microflow_diagram}b, Index-1 (step~\circledtext{2b}) and Index-2 (step~\circledtext{3b}) which correspond to entry~0 and entry~3 do not have space available. 
Therefore, the microflow cache compares the hash values of the two entries (step~\circledtext{4b}), and replaces entry 3, which has a smaller hash value than entry~0. 



\subsubsection{Megaflow Cache} \label{subsubsec:megaflow}
The microflow cache may suffer from a high miss rate in case of massive short-lived connections due to its small capacity. 
Therefore, OVS integrates a second-level cache, the megaflow cache.  
It incorporates subtables that represent common combinations of OpenFlow tables, effectively increasing the caching capacity. 
It inserts a new flow into one of its subtables (or creates a new subtable if it does not exist) upon a megaflow miss.
The subtables and entries are created based on the OpenFlow rules.
Each subtable in the megaflow cache has unique masks applied to all header fields, and entries in a subtable share the same mask.
The megaflow cache looks up subtables in order and executes the corresponding action.
However, unlike the OpenFlow table in the slow path that walks through all tables, the megaflow cache stops lookup as soon as it finds the matching entry.
There is only one matching entry among all subtables because each megaflow cache entry is unique.

\begin{figure}
    \includegraphics[width=\linewidth]{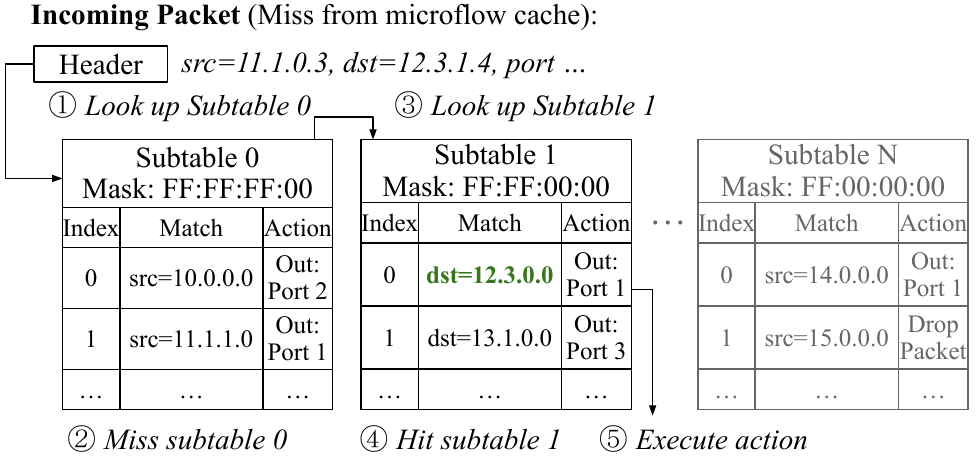}
    \caption{Megaflow cache.}
    \label{fig:Megaflow_diagram}
    
\end{figure}

\Cref{fig:Megaflow_diagram} shows an example, where the incoming packet misses the microflow cache, and is thus passed to the megaflow cache. 
The packet has source IP 11.1.0.3 and destination IP 12.3.1.4.
It first looks up subtable~0 (step~\circled{1}) but misses (step~\circled{2}). Then, it moves on to subtable 1 (step~\circled{3}) and hits after applying the bitmask (step~\circled{4}). Therefore, it skips the remaining subtables (shown in gray in \Cref{fig:Megaflow_diagram}) and executes the action from subtable~1 (step~\circled{5}). 

The size of the megaflow cache is much larger, containing a number of subtables, unlike the fixed-size microflow cache.
To reduce the lookup latency, the megaflow cache reorders subtables by periodically sorting them based on their hit counts.
Thus, frequently accessed subtables will be looked up earlier to reduce latency. 
As the megaflow cache can grow over time, it periodically evicts entries without recent hits.

\subsection{Existing Security Studies on OVS}
As OVS is intended for sharing among server tenants and attached to the network, security vulnerabilities in OVS may lead to serious consequences. 
There have been existing security studies on OVS.
Csikor \etal \cite{Csikor2020TupleSpaceOVS} perform a tuple space exploration attack on OVS that degrades OVS performance, leading to denial-of-service (DoS) for other OVS users; \u{S}raier investigates potential performance issues in OVS under untrusted network traffic~\cite{sraier2024ovssecrutiy}. 
Additionally, several potential DoS vulnerabilities in OVS have been documented in the CVE system \cite{CVE-2017-9263,CVE-2019-25076,CVE-2020-35498,CVE-2023-3966}.
However, these studies only focus on serious performance degradation caused by attackers.

The question we aim to answer is whether an attacker can break the isolation from virtualization, and acquire and transmit secret information via OVS.  
Prior studies have shown that shared caches may lead to side channels. 
As OVS contains a cache hierarchy, much like caches for databases and CPUs, in this work, we assess OVS from a side-channel security angle.

\section{OVS Characterization}

In this section, we first characterize OVS from a security angle and summarize three attack primitives in OVS.

\subsection{Experiment System}
\label{subsec:exp_setup}

\begin{table}[t]
    \caption{System configuration for OVS characterization.}
    \label{tab:system}
    \centering
    \small
    \begin{tabular}{ll}
    \toprule
    CPU  &  24 cores, Intel Xeon Cascade Lake  \\
    Memory & \SI{96}{\giga\byte} \\
    NIC & Intel I210, 1 Gbps \\
    Switch & NETGEAR GS308, 1 Gbps \\ 
    \midrule
    OS & Ubuntu 22.04, Linux kernel v6.5.0 \\
    OVS  &  OVS (v2.17.9) with DPDK datapath (v21.11.9)  \\
    \bottomrule
    \end{tabular}
\end{table}

\begin{figure}
  \centering
  \includegraphics[width=1\linewidth]{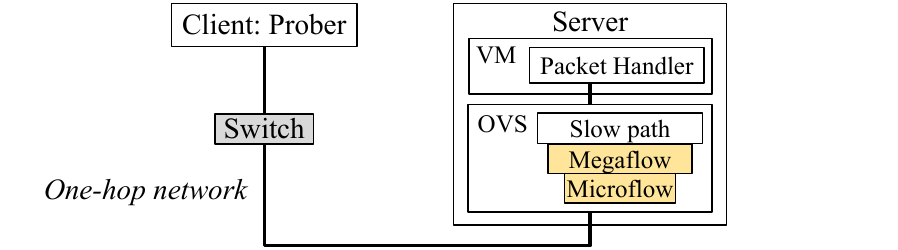}
  \caption{Overall system setup for OVS characterization.}
  \label{fig:Diagram_latency}
  
\end{figure}

\textbf{Test Platform.}
The testbed consists of two machines, one client and one server, connected by a one-hop network via a switch.
\Cref{tab:system} lists the system configuration that consists of off-the-shelf platforms and software components. 
\Cref{fig:Diagram_latency} illustrates the experiment setup.
The server hosts a virtual machine (VM) that runs a packet handler.
A virtual NIC in the VM is connected to OVS and then connects to the physical NIC on the host server. 
The prober for characterization is located on the client.
We use a Memcached instance as the handler to process \texttt{GET} requests from the prober.
This testbed is isolated to avoid interference with other users. 
We follow a proof-of-concept setup to demonstrate new side-channel vulnerabilities similar to prior remote \cite{kurth2020netcat,liu2023side,schwarz2019netspectre,schwarzl2022remote} and local \cite{gras2018translation,guo2022adversarial,liu2015last,naghibijouybari2018rendered,tan2021invisible} side-channel attacks.

\textbf{OVS Configuration.}
The OVS system uses the DPDK library as the data path \cite{ovs_dpdk}.
Its cache design has not changed since v2.6, with the microflow and megaflow caches in the current version (v3.6, December 2025) still exhibiting the same behavior discussed in \Cref{subsec:ovs}.
Another OVS version based on the kernel datapath does not include a microflow cache but still has a megaflow cache, which implications will be discussed in \Cref{subsec:attack_primitive_summary}.
We evaluate both the default microflow cache with 8k entries and a version with 16k entries by scaling the number of bits in the microflow index. 
We deploy OVS rules from ClassBench-ng \cite{classbenchng2017}, a packet classification benchmark commonly used in prior work \cite{wan2020tcache,wan2021adaptive,Stoenescu2018debuggingp4,Miano2021ebpf}.
These rules form 40 subtables in the megaflow cache. 
We follow the same configuration in the rest of the paper unless specified.

\subsection{Timing of OVS}
\label{subsec:timing_of_ovs}

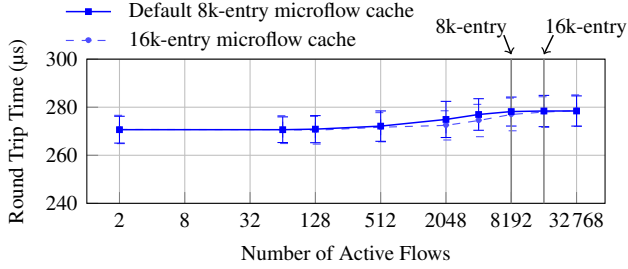
\begin{figure}
    \centering
    \begin{tikzpicture}
\begin{axis}[
reverse legend,
style={font=\footnotesize},
xlabel={Number of Active Flows},
ylabel={Round Trip Time (\textmu s)},
width=1\hsize,
scaled y ticks=false,
xtick pos=bottom,
ytick pos=left,
xtick = {2,8,32,128,512,2048,8192,32768},
xticklabels = {2,8,32,128,512,2048,8192,\SIx{32768}},
xmode=log,
xmin=1,
xmax=60000,
ymin=240,
ymax=300,
height=3.5cm,
grid=both,
legend cell align=left,
legend columns=1,
legend style={
cells={align=left},
anchor=north,
at={(0.3,1.47)},
draw=none,
fill=none,
column sep=1ex,
clip=false
},
]
\addplot+[
  white!30!blue, mark options={white!30!blue, scale=0.5},
  line width=0.15mm,
  dashed,
  error bars/.cd, 
    y fixed,
    y dir=both,
    y explicit
] table[x=x, y=y, y error=error,col sep=comma] {figures/csv/Microflow_unique_16k.csv};
\addplot+[
  line width=0.2mm,
  blue, mark options={blue, scale=0.5},
  error bars/.cd, 
    y fixed,
    y dir=both,
    y explicit
] table[x=x, y=y, y error=error,col sep=comma] {figures/csv/Microflow_unique.csv};
\draw[line width=0.2mm, color=gray] (8192,240) -- (8192,300);
\node[draw=none,color=black] at (3500,313) {8k-entry};
\draw[->, line width=0.2mm, color=black] (20000,310) -- (16384,302);

\draw[line width=0.2mm, color=gray] (16384,240) -- (16384,300);
\node[draw=none,color=black] at (40000,313) {16k-entry};
\draw[->, line width=0.2mm, color=black] (7000,310) -- (8192,302);

\legend{16k-entry microflow cache, Default 8k-entry microflow cache }

\end{axis}
\end{tikzpicture}
    \caption{Network RTT of OVS vs. number of active flows.}
    \label{fig:ovs_timing}
\end{figure}

To understand the network timing of OVS, we first measure the network round trip time (RTT) of OVS by varying the number of active flows (\SIx{100} times for each). 
Each round-trip measurement involves a \emph{forward} packet from the prober client and a \emph{backward} response packet, doubling the usage of flow entries in OVS caches. 
\Cref{fig:ovs_timing} shows the result, where the number of active flows (x-axis) counts both forward and backward directions, and each point is an average RTT (y-axis) with an error bar, representing one standard error ($\pm \sigma$).\footnote{Error bars in the remainder of this paper follow the same format.}

We first evaluate the default 8k-entry microflow cache size. 
We observe that when the number of active flows is low, the latency is the lowest ($\text{avg}=\SI{270.65}{\micro\second}$, $\sigma=2.0\,\%$). 
The latency starts to increase when the number of flows reaches the size of the microflow cache (default 8k), where the average latency is \SI{277.18}{\micro\second} ($\sigma=1.9\,\%$). 
With more flows, the latency is about the same, which is the megaflow cache hit latency.
We also evaluate a larger microflow cache with 16k entries.
As \Cref{fig:ovs_timing} shows, the microflow cache hit/miss latencies remain almost the same as the default 8k-entry setup, where the hit latency is {\SI{270.74}{\micro\second}} ($\sigma=2.1\,\%$). 

\begin{figure}
    \begin{subfigure}[b]{1\linewidth}
      \centering
      \begin{tikzpicture}
\begin{axis}[
width=1\hsize,
style={font=\footnotesize},
hide axis,         
xmin=10,
xmax=50,
ymin=0,
ymax=0.4,
legend cell align=left,
legend columns=2,
legend style={
cells={align=left},
anchor=north,
at={(0.5,1)},
draw=none,
fill=none,
column sep=1ex,
},
legend image post style={scale=0.7}, 
legend image code/.code={%
    \draw[#1,draw] (0cm,-0.03cm) rectangle (0.1cm,0.12cm);
}
]
\addlegendimage{area legend, fill=blue, draw=blue, fill opacity=0.3, draw opacity=0.3}
\addlegendentry{Microflow Hit (8k-entry)}
\addlegendimage{area legend, fill=white!60!blue, draw=white!30!blue,fill opacity=0.3, dashed}
\addlegendentry{Microflow Hit (16k-entry)}
\addlegendimage{area legend, fill=red, draw=red, fill opacity=0.3}
\addlegendentry{Megaflow Hit}
\addlegendimage{area legend, fill=white!40!gray, draw=black}
\addlegendentry{Slow Path}

\end{axis}
\end{tikzpicture}
    \end{subfigure}
    \begin{subfigure}[t]{0.55\linewidth}
      \centering
      \begin{tikzpicture}
\begin{axis}[
ybar,
style={font=\footnotesize},
xlabel={Round Trip Time (\textmu s)},
ylabel={Frequency},
width=1\hsize,
scaled y ticks=false,
xtick pos=bottom,
ytick pos=left,
ytick = {0,0.05,0.1,0.15,0.2},
yticklabels={0\,\%, 5\,\%, 10\,\%, 15\,\%, 20\,\%},
xmin=250,
xmax=300,
ymin=0,
ymax=0.2,
clip=false,
height=3.5cm,
grid=both,
legend cell align=left,
legend columns=3,
legend style={
cells={align=left},
anchor=north,
at={(0.45,1.42)},
draw=none,
fill=none,
column sep=0.5ex,
},
legend image code/.code={%
    \draw[#1,draw] (0cm,-0.04cm) rectangle (0.12cm,0.12cm);
}
]
\addplot+[
    fill = blue, draw=blue, fill opacity=0.3, draw opacity=0.3,
    hist={bins=30,data min=250, data max=300},
    y filter/.expression={y/100},
] table[y index=0] {figures/csv/cache_histogram/microflow_histogram.csv};
\addplot+[
    fill=white!60!blue, draw=white!30!blue,fill opacity=0.3, dashed,
    hist={bins=30,data min=250, data max=300},
    y filter/.expression={y/100},
] table[y index=0] {figures/csv/cache_histogram/microflow_histogram_16k.csv};
\addplot+[
    fill=red, draw=red, fill opacity=0.3,
    hist={bins=30,data min=250, data max=300},
    y filter/.expression={y/100},
] table[y index=0] {figures/csv/cache_histogram/megaflow_histogram.csv};

\draw[color=black]  (250,0) -- (300,0);

\node[draw=none, color=blue, fill=none, inner sep=0.5mm] at (261,0.18) {270 \textmu s};
\node[draw=none, color=red, fill=none, inner sep=0.5mm] at (287,0.18) {277 \textmu s};
\draw[dashed, line width=0.5mm, color=white!60!blue] (270,0) -- (270,0.2);
\draw[dashed, line width=0.2mm, color=blue] (270,0) -- (270,0.2);
\draw[dashed, line width=0.3mm, color=red] (277.18,0) -- (277.18,0.2);

\end{axis}
\end{tikzpicture}
    \end{subfigure}
    \begin{subfigure}[t]{0.44\linewidth}
      \centering
      \begin{tikzpicture}
\begin{axis}[
ybar,
style={font=\footnotesize},
xlabel={Round Trip Time (\textmu s)},
width=1\hsize,
scaled y ticks=false,
xtick pos=bottom,
ytick pos=left,
ytick = {0,0.05,0.1,0.15,0.2},
yticklabels={0\,\%, 5\,\%, 10\,\%, 15\,\%,20\,\%},
xmin=1050,
xmax=1350,
ymin=0,
ymax=0.2,
clip=false,
height=3.5cm,
grid=both,
legend cell align=left,
legend columns=1,
legend style={
cells={align=left},
anchor=north,
at={(0.15,1.42)},
draw=none,
fill=none,
column sep=0.5ex,
},
legend image code/.code={%
    \draw[#1,draw] (0cm,-0.04cm) rectangle (0.12cm,0.12cm);
}
]
\addplot+[
    hist={bins=25,data min=1050, data max=1350},
    y filter/.expression={y/100},
    fill=white!40!gray,
    draw=black,
] table[y index=0] {figures/csv/cache_histogram/slowpath_histogram.csv};

\node[draw=none, color=black, fill=none, inner sep=0.5mm] at (1122,0.18) {1196 \textmu s};
\draw[dashed, line width=0.3mm, color=black] (1196,0) -- (1196,0.2);

\end{axis}
\end{tikzpicture}
    \end{subfigure}
\caption{Latency distributions of microflow (8k- and 16k-entry have overlapped distribution), megaflow, and slow path. }
\label{fig:Cache_histogram}
\end{figure}
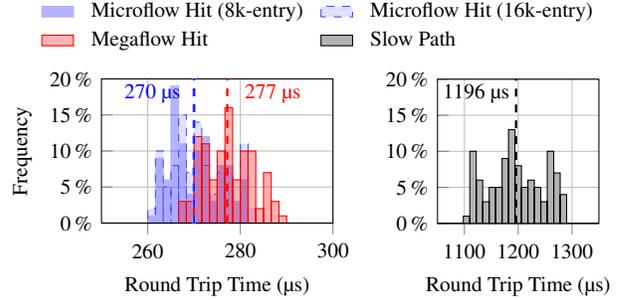

\Cref{fig:ovs_timing} does not include the slow path latency because the megaflow cache does not have a size limit.
Instead, we measure the first occurrence of a flow as the slow path latency, which has an average latency of \SI{1196.03}{\micro\second} ($\sigma=4.3\,\%$).
\Cref{fig:Cache_histogram} (left) summarizes the RTT distributions of microflow cache hits and megaflow cache hits, showing partially overlapped distributions but still distinguishable with multiple measurements as the average values differ. 
\Cref{fig:Cache_histogram} (right) shows the RTT distribution of the slow path, which is distinct from the microflow and megaflow hits. 

\textbf{Conclusion.}
The microflow cache, megaflow cache, and the slow path have distinguishable latencies. 
Repeated measurements enable more reliable differentiation of these latency levels.
Specifically, the microflow cache does exact-match, which indicates that observing a microflow cache hit latency guarantees that no other  flows use the same cache entry; a megaflow cache hit latency (a microflow miss) indicates that there has been a collision. 
In comparison, the megaflow cache maintains flows with the same header mask in the same subtable. Therefore, a megaflow hit latency indicates that the subtable is shared with other flows. 

\subsection{Microflow Cache}
\label{subsection:microflow_cache_profile}
We characterize the microflow cache by evaluating the default 8k-entry setup. 
Because its size does not change its logic or policy, conclusions drawn from these experiments remain the same when scaling the microflow cache size. 
Then, we summarize the attack primitives based on our findings.

\subsubsection{Microflow Cache Timeout}
As discussed in \Cref{subsubsec:microflow}, the microflow cache only does eviction upon collision.
We confirm whether the microflow cache ent ries time out in this experiment.
First, the client creates an entry (\ie a flow) in the microflow cache by sending an arbitrary packet. Then, we measure the RTT of the same flow after different waiting times by resending the packet. We repeat this experiment \SIx{100} times under different waiting time intervals. 
As \Cref{fig:microflow_timeout} shows, the RTT remains almost the same ($\text{avg}=\SI{270}{\micro\second}$, $\sigma=1.4\,\%$) even after \SI{100}{\second}. Therefore, the microflow cache entries do not time out. 

\begin{figure}
\begin{minipage}[b]{0.41\linewidth}
    \centering
    \begin{tikzpicture}
\begin{axis}[
style={font=\footnotesize},
xlabel style={align=center},
xlabel={Time between Two Packets\\to the Same Flow Entry (s)},
ylabel style={align=center},
ylabel={Round Trip Time (\textmu s)},
ylabel shift={-2pt},
width=1\hsize,
scaled y ticks=false,
xtick pos=bottom,
ytick pos=left,
xmax=150,
xmode=log,
ymin=100,
ymax=400,
height=3.8cm,
grid=major,
legend cell align=left,
]
\addplot+[
  black, mark options={blue, scale=0.5},
  error bars/.cd, 
    y fixed,
    y dir=both,
    y explicit
] table[x=x, y=y, y error=err,col sep=comma] {figures/csv/Microflow_timeout.csv};
\end{axis}
\end{tikzpicture}
    \caption{Microflow cache hit latency vs. waiting time ($n=100$ per data point, 8k-entry microflow cache). }
    \label{fig:microflow_timeout}
\end{minipage}
\hspace{1mm}
\begin{minipage}[b]{0.55\linewidth}
  \centering
  \begin{tikzpicture}
\begin{axis}[
ybar,
bar width=250,
style={font=\footnotesize},
xlabel={Microflow Entry Num (0 -- 8191)},
ylabel={Eviction Probability},
width=1\hsize,
scaled y ticks=false,
xtick pos=bottom,
ytick pos=left,
ytick = {0,0.1,0.2,0.3},
yticklabels={0\,\%, 10\,\%, 20\,\%, 30\,\%},
ylabel shift={-2pt},
xmin=0,
xtick = {0, 1038, 1951, 4095, 5418, 6639, 8191},
xmax=8192,
ymin=0,
ymax=0.3,
xticklabel style={rotate=90},
height=3.4cm,
ymajorgrids=true,
reverse legend,
legend columns=3,
legend cell align=left,
column sep=0.5,
legend image post style={scale=0.75}, 
legend style={nodes={scale=1},draw=none,anchor=north,at={(0.4,1.35)}, fill=none},
legend image code/.code={%
    \draw[#1,draw] (0cm,-0.04cm) rectangle (0.12cm,0.12cm);
}
]
\addplot+[
    draw=none,
    fill=gray,
    y filter/.expression={y/80},
] table[x=x,y=val,col sep=comma] {figures/csv/Microflow_header.csv};
\addlegendimage{draw=gray, fill=gray};
\legend{Noise}

\draw[draw=blue, fill=blue!40] (1038-125,0) rectangle (1038+125,20/80);
\addlegendimage{draw=blue, fill=blue!40};
\addlegendentry{Backward}
\draw[draw=blue, fill=blue!40] (6639-125,0) rectangle (6639+125,19/80);
\draw[draw=black!20!orange, fill=orange!40] (1951-125,0) rectangle (1951+125,20/80);
\draw[draw=black!20!orange, fill=orange!40] (5418-125,0) rectangle (5418+125,20/80);
\addlegendimage{draw=black!20!orange, fill=orange!40};
\addlegendentry{Forward}


\end{axis}
\end{tikzpicture}
  \caption{Microflow cache eviction (default 8k entries). Eviction packet is first sent to server (\emph{forward}) and then triggers response (\emph{backward}).}
  \label{fig:Microflow_header}
\end{minipage}
\end{figure}

\subsubsection{Collisions in Microflow Cache} \label{subsubsec:microflow_collision}

The microflow cache evicts by collision upon new flows, as introduced in \Cref{subsubsec:microflow}.
An evicted flow will be a miss if accessed again, leading to longer latency
(as shown in \Cref{fig:Cache_histogram}).
%
%
%
%
%
We investigate its behavior using the setup in \Cref{subsec:exp_setup}.
The prober periodically measures the latencies of all 8k microflow cache entries on the host.
Also, the prober sends an additional eviction packet to the packet handler on the server in between every 8k RTT measurements to generate microflow cache evictions. 
Like the experiment in \Cref{subsec:timing_of_ovs}, the additional packet also leads to two flows: forward and backward.
As discussed in \Cref{subsubsec:microflow}, each flow triggered by a network packet looks up two microflow cache entries and replaces one of them according to their hash values.
Therefore, to demonstrate evictions of both hash indices, we control the hash value of the probing packets by setting the IP address and port number.
First, the prober generates packets to probe all microflow entries with the same header fields for 100 times for higher accuracy and chooses 2 entries with the highest average RTT.
Then, for the 2 chosen entries that collide with the eviction packets, the prober modifies their IP and port number to get higher hash values than any existing hash values in the microflow cache.
This way, these entries will not be evicted, but instead have their other correspondent indices evicted. Again, the prober performs probing with new hash values for another 100 times and chooses 2 entries with the highest average RTT.
We evaluate this 20 times and summarize the probability of evictions among \SIx{8192} microflow cache entries in \Cref{fig:Microflow_header}. 
Because the prober generates evictions on both entry indices, we observe four peaks in total, where two correspond to the forward flow and two correspond to the backward flow;
other entries are evicted at a low probability ($<1.3\,\%$) due to noise.

\textbf{Conclusion.}
The hash value that indexes the microflow cache is generated from the network packet header.
Thus, entries with identical hash values may lead to a collision. 

\subsection{Megaflow cache}
\label{subsection:megaflow_cache_profile}
In this experiment, we characterize the megaflow cache. 
As it is accessed only upon microflow cache misses, all experiments below generate additional traffic to the packet handler to evict the microflow cache entry (based on colliding hash values), prior to megaflow cache measurements. 

\begin{figure}
  \centering
  \begin{tikzpicture}
\begin{axis}[
style={font=\footnotesize},
xlabel={Time between Two Packets to Same Flow Entry (s)},
ylabel={Round Trip Time (\textmu s)},
width=1\hsize,
scaled y ticks=false,
xtick pos=bottom,
ytick pos=left,
xmin=0,
xmax=16,
ymin=0,
ymax=1500,
height=3.5cm,
grid=both,
legend cell align=left,
]
\addplot+[
  black, mark options={blue, scale=0.5},
  error bars/.cd, 
    y fixed,
    y dir=both,
    y explicit
] table[x=x, y=y, y error=err,col sep=comma] {figures/csv/Megaflow_timeout.csv};
\node[draw=none, color=black!40!green, fill=white, inner sep=0.5mm] at (3,700) {Megaflow Cache Hit};
\draw[->, line width=0.3mm, color=black!40!green]  (5,600) -- (6,350);
\node[draw=none, color=black!10!red, fill=white, inner sep=0.5mm] at (5,1200) {Megaflow Cache Starts to Miss};
\draw[->, line width=0.3mm, color=black!10!red]  (6,1100) -- (9.5,600);
\end{axis}
\end{tikzpicture}
  \caption{Megaflow timeout ($n=100$ for each data point).}
  \label{fig:Megaflow_timeout}
  
\end{figure}
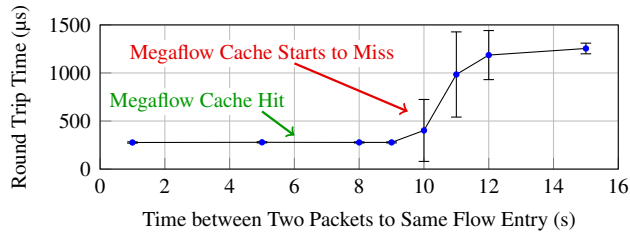

\subsubsection{Megaflow Cache Timeout} \label{subsubsec:megaflow_timeout}
Unlike the microflow cache, the megaflow cache periodically removes unused entries in subtables.
As introduced in \Cref{subsubsec:megaflow}, the minimum timeout eviction interval is \SIx{10} seconds.
In this experiment, we confirm this mechanism with a timeout eviction test on the megaflow cache. 
\Cref{fig:Megaflow_timeout} demonstrates RTT before and after timeout eviction, depending on the time between two consecutive packets.
The prober first sends a packet to the packet handler on the server and starts waiting for a time interval (x-axis).
Note that to eliminate microflow cache hits, the prober immediately evicts this packet flow from the microflow cache by sending a different but colliding packet. 
After the intended time interval has elapsed, the prober sends the second packet to the server and measures its RTT (y-axis) --- a megaflow hit latency indicates the previous entry has not timed out, and a longer latency indicates the timeout has happened. 
For each time interval, we repeat the experiment 100 times. 
We observe that the megaflow cache starts to make timeout evictions after \SI{10}{\second}, as RTT increases at that point. 
When evictions start to happen, the standard deviation is high (as indicated by the error bars).





\begin{figure}
\begin{minipage}[b]{0.52\linewidth}
  \centering
  \begin{tikzpicture}
\begin{axis}[
style={font=\footnotesize},
xlabel style={align=center},
xlabel={Packet Rate to\\Target Subtable (packet / s)},
ylabel={Round Trip Time (\textmu s)},
ylabel shift={-2pt},
width=1\hsize,
scaled y ticks=false,
xtick pos=bottom,
ytick pos=left,
xmax=1500,
xmode=log,
ymin=240,
ymax=320,
height=3.5cm,
grid=major,
legend cell align=left,
]
\addplot+[
  black, mark options={blue, scale=0.6},
  error bars/.cd, 
    y fixed,
    y dir=both,
    y explicit
] table[x=x, y=y, y error=err,col sep=comma] {figures/csv/Megaflow_target.csv};
\end{axis}
\end{tikzpicture}
  \caption{Megaflow cache reorder triggered by probing ($n=100$ per data point).}
  \label{fig:Megaflow_target}
\end{minipage}
\hspace{2mm}
\begin{minipage}[b]{0.44\linewidth}
  \centering
  \begin{tikzpicture}
\begin{axis}[
ybar,
style={font=\footnotesize},
xlabel={Reorder Interval (s)},
ylabel={Frequency},
width=1\hsize,
scaled y ticks=false,
xtick pos=bottom,
ytick pos=left,
ytick = {0,0.1,0.2,0.3,0.4},
yticklabels={0\,\%,10\,\%,20\,\%,30\,\%,40\,\%},
ylabel shift={-2pt},
xmin=0.900,
xmax=1.500,
ymin=0,
ymax=0.4,
height=3.8cm,
grid=both,
legend cell align=left,
]
\addplot+[
    hist={bins=30,data min=900, data max=1500},
    y filter/.expression={y/100},
    x filter/.expression={x/1000}
] table[y index=0] {figures/csv/Megaflow_interval.csv};
\node[draw=none, color=black!10!red, fill=white, inner sep=0.5mm, align=center] at (1.240,0.3) {Reorder\\happens\\after \SI{1}{\second}};
\draw[dashed, line width=0.6mm, color=black!10!red] (1.000,0) -- (1.000,0.4);
\draw[->, line width=0.3mm, color=black!10!red]  (1.090,0.35) -- (1.010,0.35);
\end{axis}
\end{tikzpicture}
  \caption{Subtable reorder intervals in megaflow cache ($n=100$).}
  \label{fig:Megaflow_interval}
\end{minipage}
\end{figure}

\subsubsection{Subtable Reordering}
\label{subsubsec:megaflow_subtable_reordering}

As introduced in \Cref{subsubsec:megaflow}, the order of subtables depends on their hit counts. We characterize this correlation using the following approach. 
First, we control the background traffic that accesses 30 out of 40 subtables at fixed rates (from 1 to \SI{10}{\packet\per\second} on different subtables) to keep the same order among them.
We follow the same approach in \Cref{subsubsec:megaflow_timeout} to clear microflow cache entries. 
Then, the prober targets one of the remaining 10 subtables and alters the traffic intensity from 1 to \SI{1000}{\packet\per\second}. 
\Cref{fig:Megaflow_target} shows the RTT of the target subtable --- the RTT reduces as the traffic intensity increases. 
The RTT stays around \SI{291}{\micro\second} when the prober sends \SI{1}{\packet\per\second} and reduces to \SI{277}{\micro\second} as the prober hits the megaflow cache at a higher rate.
This experiment confirms that the megaflow subtables are sorted by their hit counts.

We also investigate the time to reorder after the access count changes.
\Cref{fig:Megaflow_interval} shows the distribution of the subtable reorder interval.
We find that the minimum reorder interval is \SI{1}{\second} and 86\,\% of the intervals are within \SI{1.2}{\second}.
There is a tail that extends to \SI{1.43}{\second}, due to software variations.
For example, OVS defers reordering if threads are busy with other tasks, such as OpenFlow rule insertion and slow path access.

\textbf{Conclusion.}
The megaflow cache consists of a number of subtables. Subtables ordered to the front will be accessed first and have lower latencies. 
The megaflow cache periodically (every 1 -- \SI{1.43}{\second}) sorts subtables by their hit counts to reduce the latency of frequently accessed subtables. 
Thus, the access latency of a subtable is correlated with its access frequency. 
Moreover, entries in the subtables get evicted after \SI{10}{\second} if they are not hit by future packets.

\subsection{Summary of Attack Primitives in OVS}  \label{subsec:attack_primitive_summary}

Based on the finding, we have identified three attack primitives in OVS.

\begin{enumerate}[leftmargin=*, nolistsep]
    \item[\textbf{1.}] \textbf{Timing differences among microflow cache, megaflow cache, and slow path} (\Cref{subsec:timing_of_ovs}): Hits and misses on microflow and megaflow caches cause latency differences in network round trip latency. 
    Through the latency differences, an attacker can monitor the OVS cache status, further enabling them to infer network activities. 
    \item[\textbf{2.}] \textbf{Microflow cache hash collisions} (\Cref{subsubsec:microflow_collision}): The hash value that indexes the microflow cache is generated from five packet header fields. 
    A packet with an identical header can lead to hash collisions in the microflow cache, which allows an attacker to infer the packet header fields.
    \item[\textbf{3.}] \textbf{Megaflow cache subtable ordering} (\Cref{subsubsec:megaflow_subtable_reordering}): The megaflow cache periodically sorts subtables to minimize the hit latency.     
    An attacker can manipulate the subtable ordering by controlling the rate of packets that go through the target subtable. 
    Further, the \SI{10}{\second} timeout interval allows attackers to reset the megaflow cache state. 
    However, this also requires attacks based on the megaflow cache to complete before the timeout.
\end{enumerate}

\textbf{Discussion.}
As discussed in \Cref{subsec:exp_setup}, this setup assumes a simple one-hop network. 
Real-world environments may introduce additional noise due to more complex, multi-hop networks. 
We expect future work to explore the OVS attack primitives under more complex network configurations to better understand the impact of complex topologies.
This work uses the OVS version with the DPDK datapath, which has both the microflow and megaflow caches.
In comparison, the kernel version of OVS only has the megaflow cache. 
If a kernel version OVS is deployed, Attack Primitive 1 will have two levels of latency differences, Attack Primitive 2 will not apply due to the absence of the microflow cache, and Attack Primitive 3 will remain the same as the kernel version still maintains the megaflow subtables.

\section{Remote Covert Channel} \label{sec:covert}

In this section, we introduce two remote covert channels using the microflow cache and megaflow cache in OVS.

\begin{figure}[t]
  \centering
  \includegraphics[width=1\linewidth]{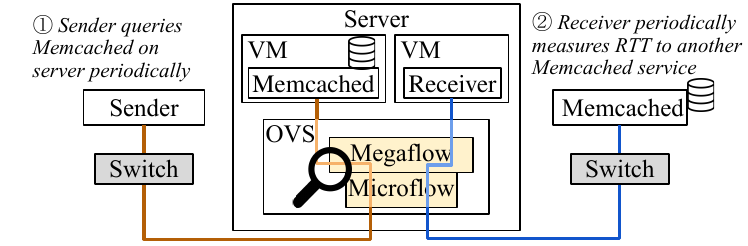}
  \caption{Setup of OVS covert channels.}
  \label{fig:Covert}
  
\end{figure}

\subsection{Attack Model}  \label{subsec:covert_model}
The sender and the receiver are connected through a one-hop network via a switch, as \Cref{fig:Covert} illustrates. However, they do not have any direct connection for communication. 
The receiver is co-located with a Memcached service on the same server, running in separate VMs and connected with the same OVS on the host server. 
The sender has access to the co-located Memcached.
The receiver also has access to another remote Memcached service that is external to its server, which is also connected through a one-hop network.
Note that we use Memcached services for demonstration. The sender and the receiver can also leverage other network services such as databases and websites. 
This attack assumes that the sender and the receiver know the microflow cache entry and megaflow subtables they will access, according to the prior knowledge of the network setup (IP, port, and protocol).
Cloud systems typically deploy a large number of OpenFlow rules, leading to numerous subtables in the megaflow cache \cite{mckeown2008openflow, rashelbach2022scaling}. Therefore, the sender and the receiver assume substantial timing differences among subtable locations.

\subsection{Attack Design}  \label{subsec:covert_design}

We detail the design of covert channel using both collisions in the microflow cache and subtables in the megaflow cache. 

\subsubsection{Microflow Cache}  \label{subsubsec:microflow_covert}
The microflow-cache-based covert channel uses the common Prime+Probe approach \cite{liu2015last}, using Attack Primitives 1 and 2.
It exploits the latency difference between a microflow cache hit and a megaflow cache hit (\ie upon microflow cache miss), as shown in \Cref{fig:Cache_histogram}.
Periodically, the sender queries the Memcached which is co-located with the receiver if it sends bit ``1'' to the receiver (step~\circled{1}), where the query takes two flow entries (forward and backward) in the microflow cache, as discussed in \Cref{subsubsec:microflow_collision}.
If the sender sends bit ``0'', it stays idle.
The receiver knows the microflow cache entries that the sender uses based on prior knowledge (as discussed in \Cref{subsec:covert_model}. 
Thus, the receiver periodically probes the same microflow cache entries (step~\circled{2}). 
Packet flows from the sender will evict the receiver's microflow cache entries. Thus, if the receiver measures a high RTT, the sender has sent bit ``1'', and otherwise, the sender has sent bit ``0''.


\subsubsection{Megaflow Cache}  
\label{subsubsec:megaflow_covert}

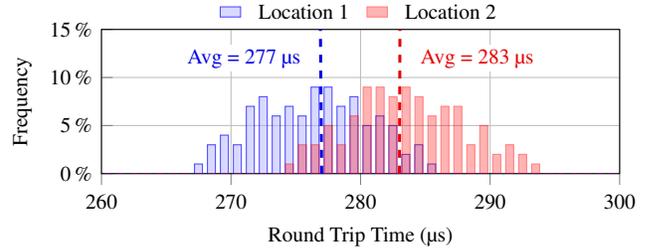
\begin{figure}
  \centering
  \begin{tikzpicture}
\begin{axis}[
ybar,
bar width=\hsize/80,
style={font=\footnotesize},
xlabel={Round Trip Time (\textmu s)},
ylabel={Frequency},
width=1\hsize,
scaled y ticks=false,
xtick pos=bottom,
ytick pos=left,
ytick = {0,5,10,15},
yticklabels={0\,\%,5\,\%,10\,\%, 15\,\%},
xmin=260,
xmax=300,
ymin=0,
ymax=15,
height=3.5cm,
grid=both,
legend columns=2,
legend style={
anchor=north,
at={(0.5,1.2)},
draw=none,
fill=none,
column sep=1ex,
},
legend image code/.code={%
    \draw[#1, draw] (0cm,-0.05cm) rectangle (0.3cm,0.1cm);
},  
legend cell align=left,
legend image post style={scale=0.9}, 
legend style={draw=none,anchor=north,at={(0.5,1.25)}},
]
\addplot+[
fill opacity=0.5,
draw opacity=0.7
] table[col sep=comma, x=latency, y=freq] {figures/csv/megaflow_covert_2/group1.csv};
\addplot+[
fill=red,
draw=red,
fill opacity=0.3,
draw opacity=0.5
] table[col sep=comma, x=latency, y=freq] {figures/csv/megaflow_covert_2/group2.csv};
\node[draw=none, color=black!10!blue, fill=white, inner sep=0.5mm] at (271,12) {Avg = 277 \textmu s};
\node[draw=none, color=black!10!red, fill=white, inner sep=0.5mm] at (289,12) {Avg = 283 \textmu s};
\draw[dashed, line width=0.4mm, color=black!10!blue] (277,0) -- (276.90,15);
\draw[dashed, line width=0.4mm, color=black!10!red] (283,0) -- (283.04,15);
\legend{Location 1, Location 2}
\end{axis}
\end{tikzpicture}
  \caption{Megaflow cache hit latency ($n=100$ per location). } \label{fig:Megaflow_histogram}
\end{figure}

The megaflow-cache-based side channel is based on Attack Primitives 1 and 3.
We take the same approach in \Cref{subsubsec:megaflow_subtable_reordering} to manipulate subtable ordering. 
\Cref{fig:Megaflow_histogram} shows the latency distributions of two subtable locations, where we measure each \SIx{100} times.
Subtable location~1 is accessed \SIx{10} times every second to move the target subtable to the front, with an average latency of \SI{276.90}{\micro\second} ($\sigma=1.6\,\%$); subtable location 2 is accessed \SIx{1} time per second, being left at the end, with an average latency of \SI{283.04}{\micro\second} ($\sigma=1.5\,\%$).
The sender first queries the Memcached service co-located with the receiver to create an initial ordering among subtables. Then, the sender targets one of the subtables and changes its location by controlling the packet rate towards this subtable (step \circled{1}). 
Note that the sender clears microflow cache entries to enable hits on megaflow subtables. 
If the receiver measures a low latency from a megaflow hit on the sender-controlled subtable, the sender has sent bit ``1''; otherwise, the bit is ``0'' (step \circled{2}).


\subsection{Setup} \label{subsec:covert_setup}
We follow the system configuration described in \Cref{subsec:exp_setup}.
On both Memcached services, a known key-value pair is stored before launching the covert channel.
Thus, the sender and the receiver can access their known key-value pair in their separate Memcached services using \texttt{GET} requests. Note that the two Memcached instances are completely independent. The purpose of the \texttt{GET} requests is to generate network traffic (for the sender) and perform timing (for the receiver). 
%
For the microflow-cache-based covert channel, we evaluate both the 8k-entry and the 16k-entry versions.
The threshold that determines microflow cache miss/hit is \SI{273.63}{\micro\second}, according to the latency distribution in \Cref{fig:Cache_histogram}. 
Because the covert channel only accesses a specific microflow cache entry, the threshold is the same for different microflow cache sizes. 
To establish a megaflow-cache-based covert channel, the receiver determines the latency thresholds using the profiling approach in \Cref{subsection:megaflow_cache_profile}.
In this experiment, we use the same OVS rules as \Cref{subsec:exp_setup}. 
Thus, the threshold that determines 2 subtable locations is \SI{279.97}{\micro\second}, following the distributions in \Cref{fig:Megaflow_histogram}. 
In both covert channels, every time, the sender transmits 2 bits ``10'' for the header and 8 bits for the payload (randomly generated in evaluation). Latency measurements are performed 50 times to transmit 1 bit for high accuracy.

\subsection{Results} \label{subsec:covert_results}
In this section, we first present the bandwidth of covert channels and then conduct a sensitivity study.

\begin{table}[t]
\caption{Covert channel bandwidth and accuracy.}
\label{tab:covert_sensitivity}
\centering
\small
\setlength{\tabcolsep}{3pt}
\begin{tabular}{lcccccc}
\toprule
\multirow{2}{*}{\textbf{Channel}} & \multirow{2}{*}{\textbf{Rep.}} & \multicolumn{1}{c}{\textbf{BW} $\mid$ \textbf{$\sigma$}} & \multicolumn{3}{c}{\textbf{Accuracy} (\%) $\mid$ \textbf{$\sigma$} (\%)} \\
& & \multicolumn{1}{c}{~(bit/s) $\mid$ (\%)} &\textbf{No Noise} &\textbf{Low} &\textbf{High}\\
\midrule
\multirow{5}{*}{\shortstack[l]{Microflow\\(8k-entry)}}& 10 & \SIx{79.0} $ \mid  \SIx{0.12} $ & \SIx{71.2} $ \mid  \SIx{0.4} $ & \SIx{71.0} $ \mid  \SIx{0.4} $ & \SIx{70.6} $ \mid  \SIx{0.4} $\\
& 30 & \SIx{26.3} $ \mid  \SIx{0.08} $ & \SIx{86.9} $ \mid  \SIx{0.3} $ & \SIx{85.6} $ \mid \SIx{0.4} $ & \SIx{85.0} $ \mid  \SIx{0.4} $\\
& 50 & \SIx{15.8} $ \mid  \SIx{0.07} $ & \SIx{96.8} $ \mid  \SIx{0.2} $  & \SIx{96.4} $ \mid  \SIx{0.3} $ & \SIx{96.4} $ \mid  \SIx{0.3} $\\
& 75 & \SIx{11.9} $ \mid  \SIx{0.05} $ & \SIx{98.1} $ \mid  \SIx{0.2} $  & \SIx{96.6} $ \mid  \SIx{0.2} $ & \SIx{96.5} $ \mid  \SIx{0.3} $\\
& 100 & ~~\SIx{7.9} $ \mid  \SIx{0.04} $ & \SIx{98.4} $ \mid  \SIx{0.2} $  & \SIx{96.8} $ \mid  \SIx{0.2} $ & \SIx{96.7} $ \mid  \SIx{0.3} $\\
\midrule
\multirow{5}{*}{\shortstack[l]{Microflow\\(16k-entry)}}& 10 & \SIx{79.0} $ \mid  \SIx{0.14} $ & \SIx{70.6} $ \mid  \SIx{0.6} $ & \SIx{69.2} $ \mid  \SIx{0.7} $ & \SIx{69.4} $ \mid  \SIx{0.6} $ \\
& 30 & \SIx{26.3} $ \mid  \SIx{0.11} $ & \SIx{86.5} $ \mid  \SIx{0.4} $ & \SIx{85.1} $ \mid  \SIx{0.5} $ & \SIx{85.6} $ \mid  \SIx{0.5} $ \\
& 50 & \SIx{15.8} $ \mid  \SIx{0.07} $ & \SIx{96.9} $ \mid  \SIx{0.3} $ & \SIx{96.0} $ \mid  \SIx{0.4} $ & \SIx{95.7} $ \mid  \SIx{0.4} $ \\
& 75 & \SIx{11.9} $ \mid  \SIx{0.05} $ & \SIx{98.0} $ \mid  \SIx{0.2} $ & \SIx{96.7} $ \mid  \SIx{0.3} $ & \SIx{96.4} $ \mid  \SIx{0.3} $ \\
& 100 & ~~\SIx{7.9} $ \mid  \SIx{0.05} $ & \SIx{98.3} $ \mid  \SIx{0.2} $ & \SIx{96.9} $ \mid  \SIx{0.3} $ & \SIx{96.8} $ \mid  \SIx{0.3} $ \\
\midrule
\multirow{5}{*}{Megaflow}& 10 & ~~\SIx{0.73} $ \mid  \SIx{0.02} $ & \SIx{63.3} $ \mid  \SIx{1.1} $  & \SIx{62.2} $ \mid  \SIx{1.3} $ & \SIx{62.1} $ \mid  \SIx{1.3} $\\
& 30 & ~~\SIx{0.72} $ \mid  \SIx{0.01} $ & \SIx{75.3} $ \mid  \SIx{0.7} $ & \SIx{71.5} $ \mid  \SIx{0.9} $ & \SIx{71.9} $ \mid  \SIx{0.9} $ \\
& 50 & ~~\SIx{0.73} $ \mid  \SIx{0.01} $ & \SIx{85.7} $ \mid  \SIx{0.6} $ & \SIx{80.0} $ \mid  \SIx{0.7} $ & \SIx{79.7} $ \mid  \SIx{0.8} $ \\
& 75 & ~~\SIx{0.73} $ \mid  \SIx{0.01} $ & \SIx{87.0} $ \mid  \SIx{0.6} $ & \SIx{85.6} $ \mid  \SIx{0.6} $ & \SIx{85.3} $ \mid  \SIx{0.6} $ \\
& 100 & ~~\SIx{0.73} $ \mid  \SIx{0.01} $ & \SIx{87.4} $ \mid  \SIx{0.6} $ & \SIx{86.6} $ \mid  \SIx{0.6} $ & \SIx{86.4} $ \mid  \SIx{0.6} $ \\
\bottomrule
\end{tabular}
\end{table}

\subsubsection{Covert Channel Bandwidth and Accuracy} 
\Cref{tab:covert_sensitivity} lists the results of the covert channels that use the microflow cache (both 8k and 16k versions) and the megaflow cache. All bandwidth measurements exclude header bits. 

\textbf{Microflow Cache.}
First, we evaluate the default microflow cache with 8k entries without noise from background traffic.
With the default 50 times of measurement, the covert channel bandwidth is \SI{15.8}{\bit\per\second}.
The accuracy is also high (96.8\,\%) as the latency of microflow cache hits and misses can be distinguished through multiple measurements, according to the characterization in \Cref{fig:Cache_histogram}. 
Then, we increase the microflow cache to 16k entries. 
We find that the bandwidth and accuracy are close to the default 8k, as the microflow-cache-based covert channel only accesses a specific entry. 
We further perform a sensitivity test to analyze the impact of measurement repetitions (from 10 to 100 measurements). 
In both 8k and 16k-entry megaflow caches, an increased number of measurement repetitions improves the accuracy, but reduces the bandwidth as more measurements take longer.

\textbf{Megaflow Cache.}
We next evaluate the megaflow cache without background noise. 
With the megaflow cache, the transmission rate is lower due to the long subtables reordering interval (\Cref{fig:Megaflow_interval}). 
With the default 50 measurement repetitions, the megaflow-based covert channel achieves \SI{0.73}{\bit/\second} bandwidth and 85.7\,\% accuracy.
We also perform a sensitivity test, showing that the accuracy also benefits from repeated measurements (87.4\,\% with 100 repetitions).
However, unlike the microflow-based covert channel, its bandwidth remains unchanged because the megaflow cache reordering interval is over \SI{1}{\second}, allowing multiple measurements within this interval. 

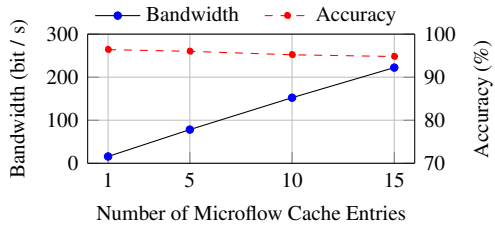
\begin{figure}
  \centering
  \begin{tikzpicture}
\pgfplotsset{set layers}

\begin{axis}[
width=.7\hsize,
style={font=\footnotesize},
xlabel={Number of Microflow Cache Entries},
ylabel={Bandwidth (bit / s)},
xtick pos=bottom,
ytick pos=left,
xmin=0,
xmax=16,
xtick={1,5,10,15},
ymin=0,
ymax=300,
axis y line*=left,
height=3.3cm,
grid=both,
legend cell align=left,
legend style={draw=none,fill=none,anchor=north,at={(0.25,1.3)}},
]
\addplot+[
  black, mark options={blue, scale=0.7},
  error bars/.cd, 
    y fixed,
    y dir=both,
    y explicit
] table[x=parallel, y=bandwidth, col sep=comma] {figures/csv/covert_parallel.csv};
\legend{Bandwidth}
\end{axis}
\begin{axis}[
width=.7\hsize,
style={font=\footnotesize},
ylabel={Accuracy (\%)},
axis x line=none,
axis y line*=right,
ytick pos=right,
xmin=0,
xmax=16,
ymin=70,
ymax=100,
height=3.3cm,
legend cell align=left,
legend style={draw=none,fill=none,anchor=north,at={(0.75,1.3)}},
]
\addplot+[
  red, mark options={red, scale=0.6}, dashed,
  error bars/.cd, 
    y fixed,
    y dir=both,
    y explicit
] table[x=parallel, y=acc, col sep=comma] {figures/csv/covert_parallel.csv};
\legend{Accuracy}
\end{axis}

\end{tikzpicture}
    \caption{Covert channel using multiple microflow entries (8k-entry microflow cache, 50 repetitions, and high noise).}
    \label{fig:covert_parallel}
\end{figure}

\subsubsection{Sensitivity Studies}
We finally conduct two sensitivity studies. 

\textbf{Noisy environment.}
We add background flows to evaluate our covert channels in a realistic environment.
We use network traffic from the UNSW-NB15 dataset \cite{moustafa2015unsw} as background noise and rank \SI{60}{\second} traces by their average packet rate. 
We choose P50 and P90 rates as the low-noise (\SI{40}{\packet\per\second}) and high-noise (\SI{150}{\packet\per\second}) scenarios, respectively. 
\Cref{tab:covert_sensitivity} also presents the accuracies of the covert channels under both noise levels.
In all covert channels, accuracy drops are less than 2\,\%.
For microflow-cache-based channels, the reason is the covert channel only targets a single entry. 
For the megaflow-cached-based channel, the microflow cache blocks most of the accesses, and thus, the noise has little impact. 

\textbf{Multiple microflow entries.}
To achieve a higher covert channel bandwidth, the sender and the receiver can use multiple microflow entries to transmit bits in parallel. 
We evaluated this approach using a configuration of 50 measurement repetitions under high-noise conditions with a standard 8k-entry microflow cache. 
\Cref{fig:covert_parallel} shows that this approach achieves \SI{222.3}{\bit\per\second} bandwidth using \SIx{15} entries without major degradation of accuracy.

\begin{table}
\caption{Comparison with other remote covert channels.}
\label{tab:existing_covert}
\centering
\setlength{\tabcolsep}{3pt}
\small
\begin{tabular}{L{4.8cm}ll}
\toprule
\textbf{Covert Channels} & \textbf{Bandwidth} & \textbf{Error} \\
\midrule
NetCAT\cite{kurth2020netcat} & \SI{16}{\kilo\bit\per\second} & \SI{0.2}{\percent} \\
Optane Persistent Memory\cite{liu2023side} & \SI{10.01}{\bit\per\second} & \SI{1.13}{\percent} \\
NetSpectre\cite{schwarz2019netspectre} & \SI{1.07}{\bit\per\second} & \SI{<0.1}{\percent} \\
Memory Deduplication\cite{schwarzl2022remote} & \SI{0.08}{\bit\per\second} & \SI{0.6}{\percent} \\
\midrule
\textbf{This work: Microflow-cache-based (using 1 microflow entry)} & \SI{15.8}{\bit\per\second} & \SI{3.2}{\percent} \\
\bottomrule
\end{tabular}
\end{table}

\subsection{Discussion}

In summary, we demonstrate covert channels using both the microflow and megaflow caches.
All these covert channels are stealthy as they are remote.

\textbf{Comparison with prior works.}
\Cref{tab:existing_covert} compares our best-performing covert channel (microflow-cache-based) with several existing remote covert channels that leverage different components in computer systems to transmit data stealthily. 
For example, Kurth \etal~\cite{kurth2020netcat} leverage Intel's data direct I/O technology (DDIO) that allows a secret channel between two remote clients, Schwarz \etal \cite{schwarz2019netspectre} implement a remote covert channel exploiting speculative execution in the server, 
Schwarzl \etal \cite{schwarzl2022remote} exploit the timing difference due to copy-on-write page faults to build a remote covert channel, and 
Liu \etal \cite{liu2023side} build a remote covert channel leveraging contentions in Optane persistent memory.
Our covert channel bandwidths and error rates are comparable with these studies.
The microflow-based covert channel achieves $1.58\times$ to $197\times$ higher bandwidth than the channels demonstrated in prior works~\cite{schwarz2019netspectre,schwarzl2022remote,liu2023side}.
Although NetCAT~\cite{kurth2020netcat} achieves higher bandwidth, their approach transmits 64 bits in parallel. As shown in \Cref{fig:covert_parallel}, it is also possible to increase the bandwidth of the microflow-based covert channel by using multiple microflow cache entries in parallel.

\textbf{Limitations.}
In our setup, OVS uses the DPDK datapath. When OVS uses the kernel datapath, there is only a megaflow cache. Consequently, the covert channel can only leverage the megaflow cache, which results in lower bandwidth.

\section{Remote Packet Header Recovery Attack} \label{sec:header_recovery}
The hash value that indexes the microflow cache is generated from five packet header fields (\Cref{subsubsec:microflow}). Therefore, it is possible to use microflow  collisions to infer the header fields. 
In particular, with prior knowledge about the service that a victim uses, it is possible to acquire some of the fields and use the microflow cache collision to infer the remaining. 
We refer to this as a \emph{packet header recovery attack}. 



\subsection{Attack Model}
\begin{figure}
    \centering
    \includegraphics[width=1\linewidth]{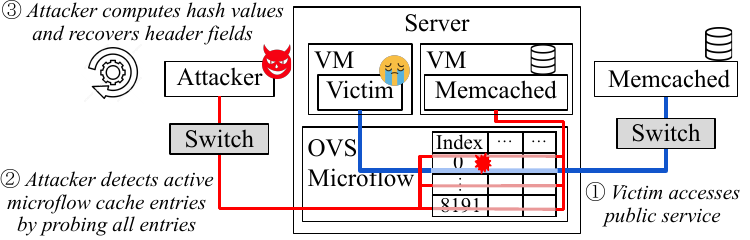}
    \caption{Setup for remote header recovery attack.}
    \label{fig:header}    
\end{figure}

\Cref{fig:header} shows the attack model, where a remote user (\ie the victim) and a Memcached service are located on the same server but isolated by separate VMs and connected with the OVS.
The attacker has access to the co-located Memcached service via queries. 
On the other hand, the victim keeps accessing a publicly accessible service, which has the IP address, port, and protocol known to the attacker.
Here, we also use Memcached queries as the victim's activity for demonstration. 
The attacker also knows the hash function used by the microflow cache, given that OVS is open-sourced. 

\subsection{Attack Design}
The header recovery attack has two stages: a \emph{probing stage} that tracks the victim's activities and an offline \emph{hash computation} stage that infers victim's header fields based on probing.

\textbf{Stage 1: Probing.}
This stage follows the Prime+Probe approach using attack primitives 1 and 2.
The victim accesses a known remote service (step \circledtext{1}).
A single packet flow looks up 2 microflow cache entries using two hash values, as discussed in \Cref{subsubsec:microflow}. 
Thus, the attacker detects both entries accessed by the victim using the approach in \Cref{subsubsec:microflow_collision}: 
the attacker first primes all microflow cache entries and probes the evicted entries (step~\circled{2}). 
To overcome noise, the attacker repeats this stage multiple times and selects the 2 entries with the highest average probing latencies. 
Then, for these 2 entries, the attacker generates probing packets with the same indices but hash values larger than existing probing packets to detect the eviction of their other associated entries, again selecting the 2 entries with the highest latency after repeated measurements.
As the victim generates one forward request packet and receives one backward response packet, the attacker finally detects a total of 4 entries evicted by the victim.


\textbf{Stage 2: Hash computation.}
With the hash values that correspond to the evicted indices known, the second stage is to figure out the source IP address and port number in the packet header that generates these hash values. 
We first demonstrate this stage using the default 8k-entry microflow cache. 
Given the microflow hashing scheme (introduced in \Cref{subsubsec:microflow}), the 32-bit source IP address and the 16-bit source port number form a total of $2^{48}$ possible pairs.
Each detected microflow cache entry reduces the number of possible pairs by $2^{13}\times$, as a 13-bit hash value decides the microflow cache entry.
In stage 1, the attacker detected 4 microflow cache entries.
Thus, one possible pair of the victim's source IP address and port number remains.
For different microflow cache sizes, as long as the size is at least $2^{12}$ entries (\ie 4k), this approach can recover an exact packet header. 
Based on this idea, the attacker computes the hash values of all possible pairs of the victim's IP address and port number using multiple cores in parallel (step \circled{3}), and finds a common pair with a hash value that matches all 4 detected entries.
This stage happens \emph{offline}, after the attacker has completed stage~1.

\subsection{Setup}
This attack uses the same one-hop network system in \Cref{subsec:covert_setup}.
The victim keeps sending \texttt{GET} requests to a remote, publicly accessible Memcached service, where the destination IP address, the destination port number, and the protocol are known and remain the same throughout the attack. 
The latency threshold for microflow cache hit/miss follows \Cref{subsec:covert_setup}.
To evaluate the effectiveness of repetition, we test stage 1 using 10, 30, 50, and 100 repetitions.


\begin{table}[t]
\caption{Time and accuracy of header recovery ($n=100$).}
\label{tab:header_recovery}
\centering
\small
\setlength{\tabcolsep}{3pt}
\begin{tabular}{lllcccc}
\toprule
\textbf{Microflow} & \textbf{Rep.} & \textbf{Time} (s) $\mid$ \textbf{$\sigma$} (\%)  &\textbf{No Noise} & \textbf{Low} &\textbf{High}  \\
\midrule
\multirow{4}{*}{8k-entry}& 10 & \quad\SIx{83.4} $\mid$ \SI{0.21}{\percent} & \SI{36}{\percent} & \SI{33}{\percent} & \SI{31}{\percent} \\
& 30 & ~~\SIx{250.8} $\mid$ \SI{0.14}{\percent} & \SI{69}{\percent} & \SI{61}{\percent} & \SI{61}{\percent}\\
& 50 & ~~\SIx{417.6} $\mid$ \SI{0.14}{\percent} & \SI{91}{\percent} & \SI{82}{\percent} & \SI{80}{\percent} \\
& 100 & ~~\SIx{837.0} $\mid$ \SI{0.12}{\percent} & \SI{93}{\percent} & \SI{87}{\percent} & \SI{87}{\percent}\\
\midrule
\multirow{4}{*}{16k-entry}& 10 & ~~\SIx{166.2} $\mid$ \SI{0.18}{\percent} & \SI{32}{\percent} & \SI{32}{\percent} & \SI{32}{\percent}\\
& 30 & ~~\SIx{501.0} $\mid$ \SI{0.14}{\percent} & \SI{70}{\percent} & \SI{59}{\percent} & \SI{60}{\percent}\\
& 50 & ~~\SIx{834.6} $\mid$ \SI{0.13}{\percent} & \SI{91}{\percent} & \SI{81}{\percent} & \SI{80}{\percent}\\
& 100 & \SIx{1673.4} $\mid$ \SI{0.09}{\percent} & \SI{94}{\percent} & \SI{86}{\percent} & \SI{85}{\percent}\\
\bottomrule
\end{tabular}
\end{table}


\subsection{Results}

\Cref{tab:header_recovery} presents the results.
We first evaluate each configuration 100 times without background noise. 
With the default 8k-entry microflow cache, attack time grows with the number of measurement repetitions in stage 1, from \SI{83.4}{\second} (10 repetitions) to \SI{837.0}{\second} (100).
The increase is due to stage 1, as stage 2 is independent of repetitions (which takes \SI{1.47}{\second}, $\sigma=\SI{1.32}{\percent}$).
Under 8k-entry microflow cache, accuracy improves with repetitions, from 36\,\% at 10 to 93\,\% at 100.
The 16k-entry cache doubles stage 1 time (twice as many entries to probe) but follows the accuracy trend, reaching 94\,\% at 100 repetitions.
As this attack targets specific microflow entries, cache size has negligible impact on accuracy.
We conclude that this attack can achieve a high accuracy with 50 or more repetitions in stage 1.

We next evaluate the attack under background noise, following the methodology in \Cref{subsec:covert_results}.
Instead of using a fixed \SI{60}{\second} noise trace, we match noise trace length with the experiment duration but still use P50 and P90 packet rates for low- and high-noise settings. 
Because packet header recovery probes the whole microflow cache, high background noise leads to an accuracy drop of 9\,\%, but still achieving 85\,\%.





\subsection{Discussion}

\textbf{Potential attacks leveraging packet header. }
An attacker can further incorporate existing activity probing attacks \cite{naghibijouybari2018rendered, tan2021invisible, lee2014stealing, shusterman2019robust, dipta2022df} or probe commonly used services to obtain the destination IP address, destination port number, and protocol since they can detect the website that the victim accesses. 
Then, the header recovery attack recovers the victim's IP address and port number.
The recovered information can be exploited by the attacker to perform other attacks, such as scanning open ports on the user's machine to attack the user's insecure services \cite{de1999review, sivanathan2018can}.  

\textbf{Comparison with prior works.}
This attack requires knowledge about the hash function that the microflow cache uses. 
Identifying the hash functions is a common approach before carrying out attacks.  
For example, Kayaalp \etal~\cite{kayaalp2016high} and Gras \etal \cite{gras2018translation} reverse-engineer hash functions in the CPU cache to let the attacker manipulate hash collisions and infer the victim's information. 
OVS uses its open-sourced hash function by default but can be configured to use other hash functions~\cite{xu2022hashing}. Therefore, prior hash reverse-engineering approaches can be applied.
Our header recovery attack computes all possible pairs of header fields. This approach is similar to the brute-force and dictionary attacks taken by prior works \cite{bovsnjak2018brute, saputra2019analysis,Owens2008ASO,javed2013sshbrute}.
In our attack, hash computation is fast and happens offline.

\textbf{Limitations.}
The attacker repeats latency measurements to achieve a reliable accuracy, which results in a long probing time. Thus, the packet header recovery attack targets long-lived flows from the victim. 
Examples include video streaming, downloading, GenAI services such as coding assistance, online meeting, online gaming, and cloud-hosted workstations. 
Moreover, if OVS employs the kernel datapath instead of the DPDK datapath, the absence of the microflow cache disables this attack. 

\section{Remote Packet Rate Monitoring Attack} \label{sec:rate}

The megaflow cache in OVS periodically reorders subtables based on packet rate, where subtables with more accesses are ordered at the front to reduce access latency as discussed in \Cref{subsubsec:megaflow}. Therefore, the ordering enables the attacker to infer the approximate rate of victim flow's packet. 
We refer to this attack as a \emph{remote packet rate monitoring attack}.


\subsection{Attack Model} \label{subsec:rate_model}
\begin{figure}[t]
  \centering
  \includegraphics[width=1\linewidth]{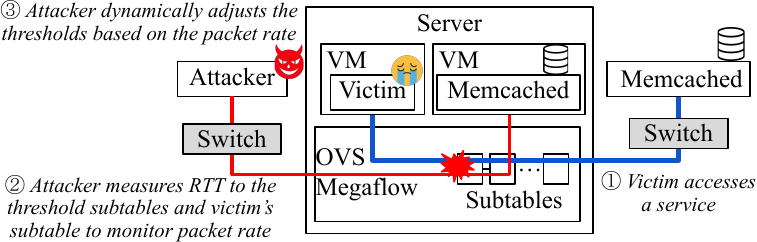}
  \caption{Setup for packet rate monitoring attack.}
  \label{fig:Rate}
  
\end{figure}

Like the packet header recovery attack, this attack targets a remote user, \ie the victim, that is co-located with a Memcached service, both running in separate VMs and connected via OVS, as shown in \Cref{fig:Rate}.
The attacker has access to this co-located Memcached using queries. 
The victim accesses a service outside the server, whose IP address, port, and protocol are known to the attacker, \eg using the packet header recovery attack in \Cref{sec:header_recovery}.
Thus, the attacker is capable of inducing cache collisions in the microflow cache to access the same megaflow subtable utilized by the victim.
We also use Memcached queries to illustrate the victim's activity and the attacker's probing access.

\subsection{Attack Design} \label{subsec:rate_design}

This attack is based on subtable reordering and the associated timing differences, leveraging Attack Primitives 1 and 3.
It is enabled by two mechanisms, packet rate monitoring and adaptive monitoring of varying packet rates.

\subsubsection{Rate monitoring} \label{subsubsec:subtable_rate_monitoring}
\begin{figure}[t]
  \centering
  \includegraphics[width=1\linewidth]{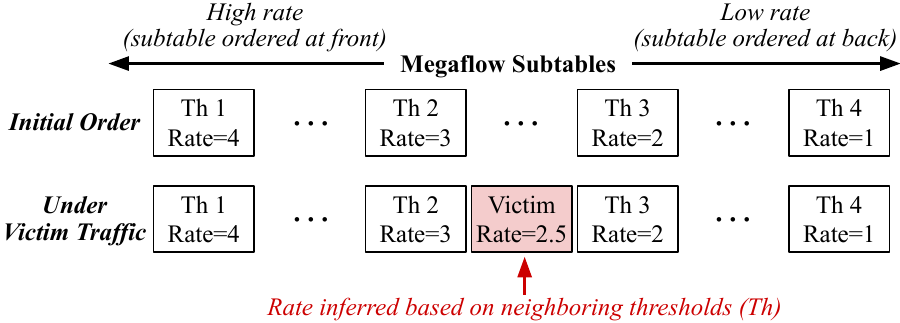}
  \caption{Subtable ordering for packet rate monitoring.}
  \label{fig:Rate_adjust}
  
\end{figure}

First, the attacker identifies megaflow subtables based on their latency differences using the same approach in \Cref{subsubsec:megaflow_subtable_reordering} and \ref{subsec:covert_model}.
While the victim is accessing a service (step~\circled{1}), the attacker periodically sends packets to 4 different subtables, where each of them receives a different packet rate.
This way, the 4 subtables are ordered and form 4 thresholds (\ie \emph{Th~1 -- Th~4} in \Cref{fig:Rate_adjust}).
\emph{Th~1} is the subtable with the highest rate of \SI{4}{\packet\per\second} and \emph{Th~4} has the lowest rate of \SI{1}{\packet\per\second}.
The attacker additionally sends the same set of rates to 5 \emph{padding subtables} in between the 4 threshold subtables to preserve distinguishable latencies between the threshold subtables.
To ensure the victim's packets end up accessing the megaflow cache, the attacker sends a packet every \SI{500}{\micro\second} to evict the victim's flow in the microflow cache, which is sufficient to support the maximum monitoring rate in \Cref{subsubsec:threshold_adjust}. 
This way, when the megaflow subtables get reordered, the subtable accessed by the victim will be ordered together with the attacker-controlled threshold subtables, as demonstrated in \Cref{fig:Rate_adjust}.
The attacker can probe the RTT of both the victim's subtable and threshold subtables and use the timing difference to infer the range of the victim's packet rate (step~\circled{2}). 
The attacker's measurement repeats \SIx{100} times every second to achieve high accuracy. 
Because probing packets also access subtables, the attacker sends extra packets at the same rate to the padding subtables to maintain the relative order.

\begin{figure}[t]
  \centering
  \includegraphics[width=1\linewidth]{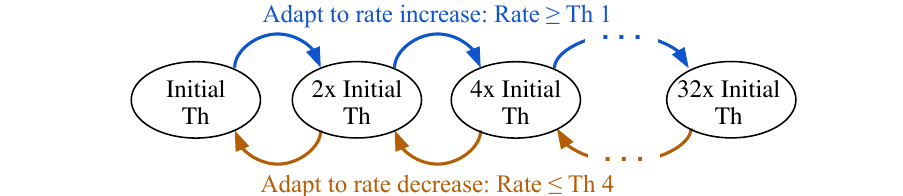}
  \caption{Dynamic threshold adjustment.}
  \label{fig:Rate_FSM}
  
\end{figure}


\subsubsection{Adaptive threshold adjustment} \label{subsubsec:threshold_adjust}
When the victim's flow has a higher packet rate than the rate of the highest threshold, the attacker can no longer track its range.
To enable tracking of varying rates, the attacker adjusts the thresholds by changing the packet rate sent to the associated megaflow subtables. 
When the victim's detected packet rate goes beyond the highest threshold (\emph{Th~1} in \Cref{fig:Rate_adjust}), the adaptive threshold adjustment mechanism doubles the thresholds' packet rate. 
The maximum thresholds in this experiment are $32\times$ of the initial values: \SIx{32}, \SIx{64}, \SIx{96}, and \SI{128}{\packet\per\second}. 
Likewise, when the monitored rate drops below the lowest threshold (\emph{Th~4}), the thresholds' packet rates are halved, until reaching the initial thresholds. 
The adjustment happens at the same rate as megaflow subtable reordering. 

\subsection{Setup} \label{subsec:rate_setup}
We follow the setup in \Cref{fig:Rate}.
The victim sends packets to a publicly accessible Memcached service, using \texttt{GET} requests.
The attacker also measures RTT to the Memcached service in the server to recover the victim's packet rate.
The OVS uses the same set of rules as \Cref{subsec:exp_setup}. 
To reduce the noise of RTT, the attacker repeats all RTT measurements 100 times.

\textbf{Dataset.} We use the UNSW-NB15 dataset \cite{moustafa2015unsw, moustafa2016evaluation} to evaluate the monitoring accuracy. 
As packet rate monitoring targets relatively long-lasting flows, we filter out short (last less than \SI{60}{\second}) and inactive flows ($<$ \SI{1}{\packet\per\second} on average).
In total, the dataset has 30 flows for the experiment. 
These flows have packet rates up to \SI{126}{\packet\per\second} and average rates between \SIx{1} and \SI{38.85}{\packet\per\second}.
We evaluate each flow for a duration of \SI{60}{\second} that captures its peak packet rate.
The victim generates packets at the rate specified by the dataset.

\textbf{Accuracy metric.}
We classify an instance as successful when the groundtruth rate falls within the same threshold as the monitored rate. Conversely, misclassification or rates falling outside the minimum and maximum thresholds are classified as unsuccessful.
We define monitoring accuracy as the ratio of successful monitoring instances over the total number of instances.

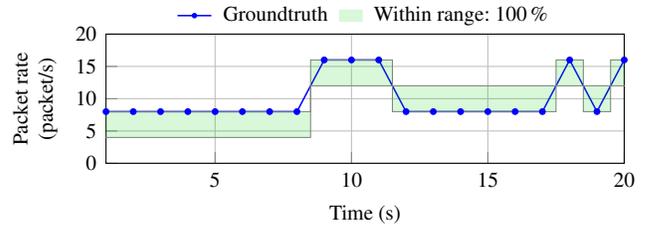
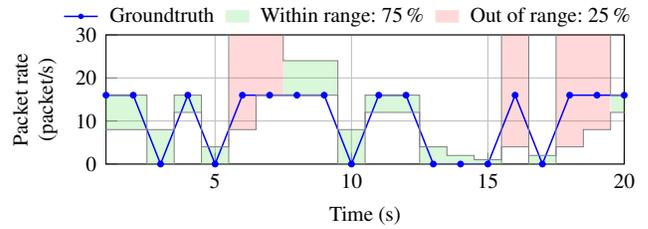
\begin{figure}[t]
\begin{subfigure}{1\linewidth}
    \centering
    \begin{tikzpicture}
\begin{axis}[
style={font=\footnotesize},
xlabel={Time (s)},
ylabel style={align=center},
ylabel={Packet rate\\(packet/s)},
width=1\hsize,
scaled y ticks=false,
xtick pos=bottom,
ytick pos=left,
xmin=1,
xmax=20,
ymin=0,
ymax=20,
height=3.3cm,
grid=both,
legend cell align=center,
legend columns=2,
legend style={
cells={align=left},
anchor=north,
at={(0.5,1.3)},
draw=none,
fill=none,
column sep=0.5ex,
clip=true
},
legend image post style={scale=0.75}, 
]
\addplot+[
  line width=0.2mm,
  blue, mark options={blue, scale=0.5},
] table[x=x, y=rate, col sep=comma] {figures/csv/rate_graph_flat2.csv};
\addplot+[
  name path=A,
  const plot,
  white!50!black, mark options={white!30!black, scale=0},
  line width=0.1mm,
  x filter/.expression={x-0.5},
  forget plot
] table[x=x, y=min, col sep=comma] {figures/csv/rate_graph_flat2.csv};
\addplot+[
  name path=B,
  const plot,
  white!50!black, mark options={white!30!black, scale=0},
  line width=0.1mm,
  x filter/.expression={x-0.5},
  forget plot
] table[x=x, y=max, col sep=comma] {figures/csv/rate_graph_flat2.csv};
\tikzfillbetween[of=A and B]{green!80!black, opacity=0.15};

\addlegendimage{legend image code/.code={\fill [green!80!black, opacity=0.15] (0cm,-0.1cm) rectangle (0.4cm,0.1cm);}}
\legend{Groundtruth, Within range: 100\,\%}

\end{axis}
\end{tikzpicture}
    \caption{Flat packet rate.}
    \label{fig:rate_graph_flat2}
    
\end{subfigure}
\begin{subfigure}{1\linewidth}
    \centering
    \begin{tikzpicture}
\begin{axis}[
style={font=\footnotesize},
xlabel={Time (s)},
ylabel style={align=center},
ylabel={Packet rate\\(packet/s)},
width=1\hsize,
scaled y ticks=false,
xtick pos=bottom,
ytick pos=left,
xmin=1,
xmax=20,
ymin=0,
ymax=30,
height=3.3cm,
grid=both,
legend cell align=center,
legend columns=3,
legend style={
cells={align=center},
anchor=north,
at={(0.47,1.3)},
draw=none,
fill=none,
column sep=0.5ex,
clip=true
},
legend image post style={scale=0.75}, 
]
\addplot+[
  line width=0.2mm,
  blue, mark options={blue, scale=0.5},
] table[x=x, y=rate, col sep=comma,] {figures/csv/rate_graph_flat.csv};
\addplot+[
  name path=A,
  const plot,
  white!50!black, mark options={scale=0},
  line width=0.1mm,
  x filter/.expression={x-0.5},
  forget plot
] table[x=x, y=min, col sep=comma] {figures/csv/rate_graph_flat.csv};
\addplot+[
  name path=B,
  const plot,
  white!50!black, mark options={scale=0},
  line width=0.1mm,
  x filter/.expression={x-0.5},
  forget plot
] table[x=x, y=max, col sep=comma] {figures/csv/rate_graph_flat.csv};
\addplot+[green!80!black, opacity=0.15, forget plot] fill between[of=A and B,split,
    every segment no 4/.style={red},
    soft clip={
        (0,-1) rectangle (7.5,1000)
    },
];
\addplot+[green!80!black, opacity=0.15, forget plot] fill between[of=A and B,split,
    every segment no 4/.style={red},
    every segment no 6/.style={red},
    soft clip={
        (7.51,-1) rectangle (19.49,1001)
    },
];
\addplot+[green!80!black, opacity=0.15, forget plot] fill between[of=A and B,split,
    soft clip={
        (19.51,-1) rectangle (21,1000)
    },
];
\addlegendimage{legend image code/.code={\fill [green!80!black, opacity=0.15] (0cm,-0.1cm) rectangle (0.4cm,0.1cm);}}
\addlegendimage{legend image code/.code={\fill [red, opacity=0.15] (0cm,-0.1cm) rectangle (0.4cm,0.1cm);}}
\legend{Groundtruth, Within range: 75\,\%, Out of range: 25\,\%}

\end{axis}
\end{tikzpicture}
    \caption{Varying packet rate.}
    \label{fig:rate_graph_flat}
    
\end{subfigure}
\caption{Demonstrations of packet rate monitoring. }
\end{figure}

\subsection{Results}

\begin{figure}[t]
\begin{subfigure}[b]{1\linewidth}
    \centering
    \begin{tikzpicture}
\begin{axis}[
ybar,
bar width=\hsize/80,
style={font=\footnotesize},
hide axis,
xlabel={Accuracy (\%)},
ylabel={Frequency},
width=1\hsize,
scaled y ticks=false,
xtick pos=bottom,
ytick pos=left,
ytick = {0,5,10,15},
yticklabels={0\,\%,5\,\%,10\,\%, 15\,\%},
xmin=0,
xmax=100,
ymin=0,
ymax=15,
height=3.2cm,
grid=both,
legend columns=4,
legend style={
anchor=north,
at={(0.5,1.2)},
draw=none,
fill=none,
column sep=0.5ex,
},
legend image code/.code={
    \draw[#1, draw] (0cm,-0.05cm) rectangle (0.1cm,0.1cm);
},  
legend image post style={scale=0.5}, 
]

\addlegendimage{empty legend}
\addlegendentry{Noise level:}
\addlegendentry{No}
\addlegendimage{area legend, draw=black, fill=black, fill opacity=0.3}
\addlegendentry{Low}
\addlegendimage{area legend, draw=blue, fill=blue, fill opacity=0.3}
\addlegendentry{High}
\addlegendimage{area legend, draw=red, fill=red, fill opacity=0.3}

\end{axis}
\end{tikzpicture}
\end{subfigure}
\begin{subfigure}[b]{0.48\linewidth}
    \centering
    \begin{tikzpicture}
\begin{axis}[
ybar,
bar width=2.5pt,
style={font=\footnotesize},
xlabel={Accuracy (\%)},
ylabel={Frequency},
width=1\hsize,
scaled y ticks=false,
xtick pos=bottom,
ytick pos=left,
ytick={0,25,50},
yticklabels = {0\,\%,25\,\%,50\,\%},
xmin=30,
xmax=100,
ymin=0,
ymax=50,
height=3.3cm,
grid=both,
legend columns=2,
legend style={
anchor=north,
at={(0.5,1.2)},
draw=none,
fill=none,
column sep=1ex,
},
]

\addplot+[
    fill=black, fill opacity=0.3, draw=black, bar shift=-2.5pt,
] table[y=no, col sep=comma] {figures/csv/rate_acc8k_hist.csv};

\addplot+[
    fill=blue, fill opacity=0.3, draw=blue,
] table[y=low, col sep=comma] {figures/csv/rate_acc8k_hist.csv};

\addplot+[
    fill=red, fill opacity=0.3, draw=red, bar shift=2.5pt,
] table[y=high, col sep=comma] {figures/csv/rate_acc8k_hist.csv};


\end{axis}
\end{tikzpicture}
    \caption{8k-entry microflow cache.}\label{subfig:rate_accuracy_8k}
\end{subfigure}
\hspace{1mm}
\begin{subfigure}[b]{0.48\linewidth}
    \centering
    \begin{tikzpicture}
\begin{axis}[
ybar,
bar width=2.5pt,
style={font=\footnotesize},
xlabel={Accuracy (\%)},
ylabel={Frequency},
width=1\hsize,
scaled y ticks=false,
xtick pos=bottom,
ytick pos=left,
ytick={0,25,50},
yticklabels = {0\,\%,25\,\%,50\,\%},
xmin=30,
xmax=100,
ymin=0,
ymax=50,
height=3.3cm,
grid=both,
legend columns=2,
legend style={
anchor=north,
at={(0.5,1.2)},
draw=none,
fill=none,
column sep=1ex,
},
]

\addplot+[
    fill=black, fill opacity=0.3, draw=black, bar shift=-2.5pt,
] table[y=no, col sep=comma] {figures/csv/rate_acc16k_hist.csv};

\addplot+[
    fill=blue, fill opacity=0.3, draw=blue,
] table[y=low, col sep=comma] {figures/csv/rate_acc16k_hist.csv};

\addplot+[
    fill=red, fill opacity=0.3, draw=red, bar shift=2.5pt,
] table[y=high, col sep=comma] {figures/csv/rate_acc16k_hist.csv};


\end{axis}
\end{tikzpicture}
    \caption{16k-entry microflow cache.}\label{subfig:rate_accuracy_16k}
\end{subfigure}
\caption{Rate monitoring accuracy.}
\label{fig:rate_accuracy}

\end{figure}
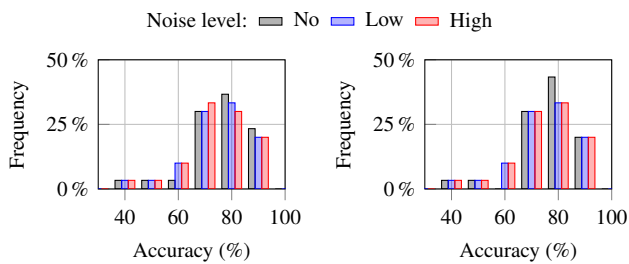

\begin{table*}[t]
    \caption{Effectiveness of defense mechanisms, \ie attack accuracies after integrating the defense mechanisms.}
    \label{tab:mitigation}
    \centering
    \small
    \begin{tabular}{llll}
    \toprule
        Attacks & OVS Isolation & Microflow Randomization & Megaflow Randomization  \\
    \midrule
        Covert channel (Microflow) & 49.60\,\% ($\sigma = 1.5\,\%$) & 53.90\,\% ($\sigma = 2.6\,\%$) & --- \\
        Covert channel (Megaflow) & 50.20\,\% ($\sigma = 1.4\,\%$) & 51.80\,\% ($\sigma = 1.8\,\%$) & 52.35\,\% ($\sigma = 2.7\,\%$) \\
        Packet header recovery & 0\,\%  & 5\,\% & --- \\
        Packet rate monitoring & 10.83\,\% ($\sigma = 6.8\,\%$) & 3.87\,\% ($\sigma = 1.4\,\%$) & 13.33\,\% ($\sigma = 6.4\,\%$) \\
    \bottomrule
    \end{tabular}
\end{table*}

\Cref{fig:rate_graph_flat2,fig:rate_graph_flat} show two \SI{20}{\second} traces.
The blue line is the victim’s ground truth; colored regions are the attacker’s monitored ranges (green for successful and red for unsuccessful).
Regions exceeding the y-axis indicate the rate exceeds the threshold.
The attack is highly accurate on flat-rate segments as the ground truth stays within the threshold ranges (\Cref{fig:rate_graph_flat2}).
When rates fluctuate, the success rate drops. For example, seconds 6--8 in  \Cref{fig:rate_graph_flat} shows a case where the victim's rate suddenly increases and the attacker's thresholds take 2 seconds to catch up.

We first evaluate each flow 20 times and record its average accuracy in an environment without background noise. 
\Cref{subfig:rate_accuracy_8k,subfig:rate_accuracy_16k} plot the accuracy distributions of the 30 flows for 8k and 16k-entry microflow caches.
Averages are \SI{71.92}{\percent} and \SI{71.83}{\percent}, respectively, with minimal sensitivity to cache size as this attack relies on megaflow cache reordering.
Spiky flows are harder to monitor, but 16.7\% of flows have over \SI{85}{\percent} accuracy.
%
We then evaluate the accuracy under noise, following the same setup in \Cref{subsec:covert_results}.
\Cref{fig:rate_accuracy} shows the accuracy of packet rate monitoring attack with the low (\SI{70.83}{\percent} and \SI{70.53}{\percent}) and high noise (\SI{70.19}{\percent} and \SI{69.81}{\percent}).
The accuracy declines by 2\,\% as background traffic is mostly blocked by the microflow cache, without affecting packet rate monitoring that uses the megaflow cache.

\subsection{Discussion}

\textbf{Potential attacks leveraging packet rates. }
The remote packet rate monitoring attack can further enable attacks such as traffic classification and activity profiling. 
The packet rate is a common feature in statistics-based network traffic classification and can be used to infer user's activities, as demonstrated by prior works \cite{zhang2014robust, erman2007semi,crotti2007traffic}. 

\textbf{Comparison with prior works.}
There are various activity profiling attacks in the literature \cite{naghibijouybari2018rendered, tan2021invisible, lee2014stealing, shusterman2019robust, qin2018website, la2021wireless, dipta2022df, yang2018usb, yang2016inferring}.
For example, Naghibijouybari \etal \cite{naghibijouybari2018rendered} leverage GPU memory utilization and Tan \etal \cite{tan2021invisible} exploit PCIe congestion due to the NIC to infer website activities.
In comparison, our attack is highly stealthy as it is remote, without requiring the attacker to be physically co-located with the victim but only accessing a service that shares OVS with the victim.
There is also a remote traffic monitoring attack \cite{liu2017flow}, which exploits timing differences between SDN rule hits and misses.
It tracks whether the victim's flow has occurred in a certain period of time to deduce the packet rate, which is a coarse-grained method. Our attack enables a real-time, more fine-grained packet rate recovery, as the attacker directly monitors the packet rate of the victim's network traffic. 

\textbf{Limitations.}
This attack achieves higher accuracy when the victim service has a stable packet rate. 
Therefore, the packet rate monitoring attack targets services that maintain a consistent packet rate over long periods of time, similar to the packet header recovery attack.
In our experimental setting that uses OVS with the DPDK datapath, the microflow cache filters out most traffic and thus enabling the attacker to monitor the megaflow cache subtables with lower noise. 
OVS with a kernel datapath can perform a similar attack. However, because it does not have a microflow cache but only a megaflow cache, the attack is more susceptible to noise.

\section{Attack Mitigations} \label{sec:defense}

In this section, we first describe three defense mechanisms that mitigate OVS side channels. Then, we discuss potential detection methods for side-channel attacks on OVS. 



\subsection{Defense Mechanisms}

We implement three defense mechanisms and evaluate their effectiveness. 
\Cref{tab:mitigation} shows the effectiveness of these mechanisms against each of the three attacks.
The evaluation follows the default configurations, where the OVS uses an 8k-entry microflow cache and the system has no background noise. 

\textbf{OVS Instance Isolation.}
The root cause of side channels in OVS is the sharing of caching structures. 
A direct approach is to eliminate this sharing by using separate instances of OVS concurrently on the server.
We evaluate this approach by deploying two separate OVS instances for the victim and the co-located service that can be accessed by the attacker. 
\Cref{tab:mitigation} shows the effectiveness of isolation. 
Both covert channels have around 50\,\% accuracy. As each bit can be either a ``0'' or ``1'' with a 50\,\% chance due to the random payload in this experiment, 50\,\% is the accuracy of a random bit stream, indicating the effectiveness of this defense.
The header recovery attack has a 0\,\% success rate, indicating that isolation eliminates this attack.
The packet rate monitoring attack has a 10.83\,\% accuracy. 
Because the isolation eliminates the victim's access to the attacker-monitored megaflow cache, almost all attacker's measurements report zero traffic (except for noise). 
As the evaluated dataset has 10.33\,\% zero traffic, this accuracy implies no actual victim traffic was successfully monitored.
However, using individual OVS can lead to extra system overheads, such as memory and CPU cycles, and complicate network management and configuration. 

\textbf{Microflow Cache Hash Randomization.}
The microflow cache uses a simple hashing scheme, where hash collisions can be used to infer user activities and steal network header information. 
A possible approach to mitigate side channels is to randomize accesses. 
For each flow, instead of colliding into two fixed entries in the microflow cache, randomized algorithms can direct the flow to more potential locations, like those used in prior cache randomization approaches \cite{wang2007new, liu2014random, qureshi2018ceaser, werner2019scattercache}. 
This mitigation can increase the candidates from hash collisions, making it harder to guess the victim's flows.
\Cref{tab:mitigation} shows the results when the randomization mechanism introduces $10\times$ candidates. 
It effectively lowers the accuracy of both types of covert channels to almost 50\,\%, demonstrating effective mitigation. 
It significantly lowers the accuracy of the header recovery attacks --- only 5 out of 100 trials by the attacker were successful. 
It also defends against the packet rate monitoring attack. Even though it uses the megaflow cache, randomization of the microflow cache disables precise microflow eviction.

\textbf{Megaflow Cache Reordering Randomization.}
The megaflow cache is another structure that can be exploited by attackers, as its subtable reordering can still be used to infer or transmit information. 
One solution is to randomize the reordering interval. Instead of performing reordering every \SI{1}{\second}, the megaflow cache can randomly perform reordering at any time in a longer interval, reducing the correlation between access frequency and subtable location. 
\Cref{tab:mitigation} shows the results when the subtable reordering interval is randomly set between 1 and \SI{10}{\second}. 
Like microflow cache hash randomization, this defense lowers the megaflow-based covert channel accuracy to almost 50\,\%, indicating effective mitigation.
It also defends against the packet rate monitoring attack. 
However, this defense is not effective against the microflow-based covert channel and the header recovery attack, as they do not rely on the megaflow cache.

\subsection{OVS Attack Detection}

An alternative approach is to detect side-channel attacks on OVS. 
These side-channel attacks mostly feature repeated accesses to a set of flows, such as performing Prime+Probe on certain flow entries in the microflow cache. 
Existing tools that support network traffic anomaly detection, such as Intrusion Detection/Prevention System (IDS/IPS)~\cite{Snort}, Network Behavior Anomaly Detection (NBAD)~\cite{Stealthwatch,Cortex_XDR}, and Security Information and Event Management (SIEM)~\cite{QRadar}, can identify suspicious network traffic introduced by the attacks.
Moreover, prior research works also suggest detection methods focused on analyzing packet patterns \cite{kim2004flow, gao2020malicious, lyamin2018ai}.
Upon detection of malicious traffic, the host may drop packets from the sender or delay these packets.
The host can also isolate packets from the potentially malicious users by directing them to a separate OVS instance. 

\section{Conclusions}
The Open vSwitch (OVS) is a widely used software-based virtual switch. 
In this work, we investigate OVS from a security perspective.
We first identify three attack primitives in the caching mechanism of OVS. 
Then, using these attack primitives, we demonstrate remote covert channels, a remote header recovery attack, and a remote packet rate monitoring attack on OVS.
Our attacks leak user information remotely, without having the attacker co-located with the victim.
Our study demonstrates that side channels via OVS can break the isolation in a virtualization environment.

\section*{Acknowledgement}

We thank the anonymous reviewers and the shepherd for their valuable feedback, and Zixuan Wang and Korakit Seemakhupt for proofreading. 
This work was supported in part by funding from the Innovation for Defence Excellence and Security (IDEaS) program from the Department of National Defence (DND).
This work was also supported by a Discovery Grant from the Natural Sciences and Engineering Research Council of Canada (NSERC).
\appendix

\section*{Ethical Considerations}

\textbf{Decision to Conduct the Research.}
We investigated Open vSwitch (OVS) due to the increasing demands of virtualized environments, where OVS serves as one of the most widely deployed virtual switches. It has been adopted across numerous platforms, including XenServer 6.0, Xen Cloud Platform, OpenStack, openQRM, OpenNebula, and oVirt.
Our study focuses on identifying potential vulnerabilities to bring this issue to the OVS developers. While vulnerability disclosure may expose potential attack vectors, we also present effective mitigation strategies and believe it is essential for stakeholders to understand both the risks and available defenses. Because OVS is extensively used in multi-tenant cloud environments, any side-channel risks that arise from resource sharing may introduce meaningful and severe security consequences.

\textbf{Stakeholders.}
The identified attacks and vulnerabilities introduce significant security risks for cloud service providers that deploy OVS, as well as for the users who rely on these environments.

\textbf{Impacts.}
As cloud environments continue to expand, virtual switches like OVS are widely deployed to improve network performance, and our findings highlight the security risks that accompany such deployments.
First, users whose traffic shares the same OVS instance as an attacker may unintentionally leak sensitive information, such as IP addresses and port numbers, as shown in our study. Second, attackers can infer users’ packet rates, potentially revealing further details about their activities. Finally, covert channels that we identify could enable unauthorized remote communication between distinct parties.
We present the potential security risks and, at the same time, propose mitigation strategies that stakeholders can adopt to prevent the proposed attacks. By characterizing the attack primitives and providing defense mechanisms, we alert users to the associated risks.

\textbf{Mitigation.}
As our proposed attacks pose severe threats to current network systems, we responsibly consider ways to mitigate these attacks.
To address the identified vulnerabilities, we propose three defense mechanisms: isolating OVS instances, randomizing the microflow cache hash, and randomizing the interval of megaflow cache reordering.
These mechanisms have been shared with the OVS developers, and we have evaluated their effectiveness. Isolation of OVS instances and randomization of the microflow cache hash successfully defend against all three attacks we propose. Randomization of the megaflow cache reordering interval mitigates megaflow-based attacks, preventing both the megaflow cache covert channel and packet-rate monitoring attacks. The corresponding results are reported in \Cref{sec:defense}.
Our defense mechanisms effectively mitigate the risks associated with sharing OVS instances on a server and prevent the attacks presented in this paper. Mitigation strategies not only prevent attacks but also protect sensitive user data, ensuring that the research contributes positively to the security of cloud environments.

\textbf{Attack Setup.}
Our evaluation was conducted entirely within an isolated network environment, as described in Section 3. To avoid any possibility of interfering with external systems or networks, all cables connecting to outside networks were physically disconnected throughout the experiments, including during the characterization phase. The servers used in the evaluation were directly interconnected, and the physical switch enabling the 1-hop network was also fully isolated, with no external connections.





\section*{Open Science}
In accordance with USENIX Security’s Open Science policy, the artifacts for this work are publicly available at \url{https://doi.org/10.5281/zenodo.17965902}.
The OpenFlow rule sets (ClassBench-ng) and the UNSW-NB15 dataset are available at \url{https://classbench-ng.github.io/} and \url{https://research.unsw.edu.au/projects/unsw-nb15-dataset}, respectively.

\bibliographystyle{plain}
\bibliography{references/misc,references/net,references/sec}

\end{document}